\documentclass[12pt]{article}
\pdfoutput=1
\usepackage{amssymb,graphicx}
\usepackage{epstopdf}
\usepackage{amsmath,amsfonts}
\usepackage{epsfig}
\usepackage{pdflscape}
\begin{document}

\title{\bf Effect of intermediate Minkowskian evolution on CMB bispectrum}
\author{S.~A.~Mironov$^{a,b}$\footnote{{\bf e-mail}: sa.mironov\_1@physics.msu.ru}, 
S.~R.~Ramazanov$^{c}$\footnote{{\bf e-mail}: Sabir.Ramazanov@ulb.ac.be},
V.~A.~Rubakov$^{a,d}$\footnote{{\bf e-mail}: rubakov@ms2.inr.ac.ru}\\
$^a$ \small{\em Institute for Nuclear Research of the Russian Academy of Sciences,}\\
\small{\em 60th October Anniversary Prospect 7a, 117312, Moscow, Russia}\\
$^b$ \small{\em Institute of Theoretical and Experimental Physics,} \\
\small{\em Bolshaya Cheremushkinskaya 25, 117218, Moscow, Russia}\\
$^c$ \small{\em Service de Physique Th\'eorique, Universit\'e Libre de Bruxelles (ULB),}\\
\small{\em CP225 Boulevard du Triomphe, B-1050 Bruxelles, Belgium}\\
$^d$ \small{\em Department of Particle Physics and Cosmology, Physics Faculty,} \\
\small{\em M.~V.~Lomonosov Moscow State University,}\\
\small{\em Vorobjevy Gory, 119991, Moscow, Russia}\\
} 
\maketitle

\begin{abstract}
We consider a non-inflationary 
early Universe scenario in which relevant scalar perturbations get frozen out 
at some point, but then are defrosted and follow a long nearly Minkowskian evolution before 
the hot era. This intermediate stage leaves specific imprint 
on the CMB 3-point function, largely independent of 
details of microscopic physics. 
In particular, the CMB bispectrum undergoes oscillations 
in the multipole $l$ space with 
roughly constant amplitude. The latter is in contrast 
to the oscillatory bispectrum enhanced 
in the flattened triangle limit, as  predicted by inflation with 
non-Bunch--Davies vacuum. 
Given this and other peculiar features of the bispectrum, 
stringent constraints imposed by the Planck data may not apply. 
The CMB 3-point function is suppressed by
the inverse duration squared of the Minkowskian evolution, but can be of 
observable size for relatively short intermediate Minkowskian stage. 
 
\end{abstract}

\section{Introduction} 

Recently released Planck data favor 
highly Gaussian primordial perturbations~\cite{Ade:2013ydc}. 
Although this is in 
agreement with predictions of the slow roll 
inflation~\cite{Maldacena:2002vr}, the 
picture of the early Universe is still uncertain. 
In particular, there remains a window for scenarios 
with relatively strong non-linearities. 
One reason is that the existing constraints  apply 
to specific types of non-Gaussianity only; 
other shapes of correlation functions require separate 
data analyses. Second, one can imagine the situation where fairly 
large non-linear effects generated during some cosmological epoch 
get diluted at later times. We entertain both of these possibilities
in this paper.

It is well known that the 
shapes of inflationary bispectra 
are strongly sensitive to the regime of evolution of
scalar perturbations at the time when the non-linearities
are at work~\cite{Babich:2004gb}. 
Conventionally, it is assumed 
that perturbations start from vacuum initial conditions, get frozen out 
at some point 
and then remain unchanged until the hot era. One then classifies possible 
non-Gaussianities as follows. 
If perturbations are superhorizon 
at the time when non-linearities are important, 
like in the curvaton~\cite{Linde:1996gt} or 
modulated reheating scenario~\cite{Dvali:2003em, Zaldarriaga:2003my}, 
then the bispectrum 
is enhanced in the squeezed limit. See Ref.~\cite{Bartolo:2004if} 
for a review. The equilateral bispectrum is characteristic of
non-Gaussianities 
generated in the on-horizon regime~\cite{Chen:2006nt}. There are also 
scenarios with inflaton fluctuations that start 
from unconventional (non-Bunch--Davies) 
initial conditions~\cite{Holman:2007na, Chialva:2011hc, Meerburg:2009ys, Agullo:2010ws, Chen:2010bka, Ashoorioon:2010xg}. 
In this situation, relevant is the evolution in the 
subhorizon regime at inflation, 
and the typical outcome is 
the oscillatory bispectrum enhanced in the flattened triangle limit.
Similar oscillatory bispectra  
originate also from the violation of the slow-roll 
conditions~\cite{Wang:1999vf} or features in the 
inflaton potential~\cite{Chen:2010bka, Flauger:2009ab}.

It may happen that cosmological perturbations
have more complicated history than  in inflationary models. 
In this paper we consider the following scenario, see Fig.~\ref{newfig}.

\unitlength 1mm 
\linethickness{0.4pt}
\ifx\plotpoint\undefined\newsavebox{\plotpoint}\fi 
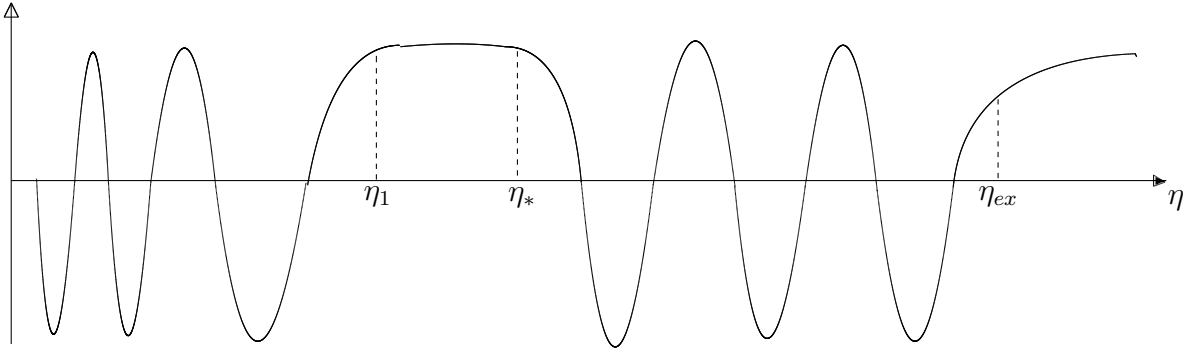
\begin{figure}
\label{evolution}
\begin{picture}(28,50)(7,50)
\setlength{\unitlength}{0.75\unitlength}
\qbezier(19,117.75)(21.625,63)(25.75,117.25)
\qbezier(25.75,117.25)(29.25,163.375)(31.75,117)
\qbezier(31.75,117)(35,62.875)(39.25,117.25)
\qbezier(39.25,117.25)(45.5,165.125)(50.75,116.5)
\qbezier(50.75,116.5)(57.75,61.25)(66.75,117)
\qbezier(142.75,116.5)(148,62.5)(155.25,116.5)
\qbezier(155.25,116.5)(162.375,166.875)(168,115.75)
\qbezier(168,115.75)(174.75,61.875)(181.5,116.5)
\qbezier(181.5,116.5)(183.875,138.75)(213.75,140)
\multiput(213.75,140)(.03125,-.0625){8}{\line(0,-1){.0625}}
\put(14.5,117.5){\vector(1,0){204.75}}
\multiput(219,117.25)(-.05263158,.03289474){38}{\line(-1,0){.05263158}}
\put(217,118.5){\line(0,-1){2}}
\multiput(217,116.5)(.0869565,.0326087){23}{\line(1,0){.0869565}}
\put(221,114){\makebox(0,0)[cc]{$\eta$}}
\put(189.43,131.93){\line(0,-1){.9333}}
\put(189.43,130.063){\line(0,-1){.9333}}
\put(189.43,128.196){\line(0,-1){.9333}}
\put(189.43,126.33){\line(0,-1){.9333}}
\put(189.43,124.463){\line(0,-1){.9333}}
\put(189.43,122.596){\line(0,-1){.9333}}
\put(189.43,120.73){\line(0,-1){.9333}}
\put(189.43,118.863){\line(0,-1){.9333}}
\put(14.5,88.5){\line(0,1){60.5}}
\multiput(14.5,149)(-.03289474,-.06578947){38}{\line(0,-1){.06578947}}
\put(13.25,146.5){\line(1,0){2.5}}
\multiput(15.75,146.5)(-.03289474,.06578947){38}{\line(0,1){.06578947}}
\put(79.5,114.75){\makebox(0,0)[cc]{$\eta_1$}}
\put(189.5,114.75){\makebox(0,0)[cc]{$\eta_{ex}$}}
\qbezier(67.05,116.75)(71.425,141.375)(83.3,141.5)
\qbezier(142.8,116.5)(136.05,167.875)(128.3,116.75)
\qbezier(128.3,116.75)(121.175,59)(115.55,117.25)
\qbezier(115.55,117.25)(113.3,141.25)(102.05,141.25)
\qbezier(102.05,141.25)(93.8,142.25)(83.55,141.25)
\put(79.23,139.629){\line(0,-1){1.1412}}
\put(79.23,137.446){\line(0,-1){.9412}}
\put(79.23,135.563){\line(0,-1){.9412}}
\put(79.23,133.68){\line(0,-1){.9412}}
\put(79.23,131.797){\line(0,-1){.9412}}
\put(79.23,129.915){\line(0,-1){.9412}}
\put(79.23,128.033){\line(0,-1){.9412}}
\put(79.23,126.15){\line(0,-1){.9412}}
\put(79.23,124.268){\line(0,-1){.9412}}
\put(79.23,122.386){\line(0,-1){.9412}}
\put(79.23,120.503){\line(0,-1){.9412}}
\put(79.23,118.621){\line(0,-1){.9412}}
\put(104.23,140.449){\line(0,-1){.9583}}
\put(104.23,138.526){\line(0,-1){.9583}}
\put(104.23,136.603){\line(0,-1){.9583}}
\put(104.23,134.68){\line(0,-1){.9583}}
\put(104.23,132.763){\line(0,-1){.9583}}
\put(104.23,130.846){\line(0,-1){.9583}}
\put(104.23,128.93){\line(0,-1){.9583}}
\put(104.23,127.013){\line(0,-1){.9583}}
\put(104.23,125.096){\line(0,-1){.9583}}
\put(104.23,123.18){\line(0,-1){.9583}}
\put(104.23,121.263){\line(0,-1){.9583}}
\put(104.23,119.346){\line(0,-1){.9583}}
\put(105.05,114.5){\makebox(0,0)[cc]{$\eta_*$}}
\end{picture}
\caption{Schematic behavior of scalar perturbations.
I. Generation epoch: $\eta< \eta_*$. Towards the end of this epoch, perturbations get frozen 
(times $\eta \sim \eta_1$) and have nearly flat power spectrum. II. Intermediate stage: $\eta_*<\eta<\eta_{ex}$. Perturbations
oscillate in nearly Minkowski space-time. III. Late stage: $\eta> \eta_{ex}$. Perturbations
are again frozen at time $\eta =\eta_{ex}$. After $\eta_{ex}$ perturbations evolve 
in the standard way (late evolution is not shown). The hot cosmological epoch starts some time after $\eta_{ex}$.
}\label{newfig}
\end{figure}
We assume that 
perturbations are generated at some early cosmological epoch, get frozen
before the end of this epoch and have nearly flat power spectrum
already at that time. Then there is an intermediate stage when 
space-time is nearly Minkowskian and perturbations are defrosted 
and oscillate. They get frozen again at the end of the intermediate stage and stay constant
until the beginning of the hot epoch and later. Their further evolution 
is standard. It is the assumed existence of the intermediate Minkowskian
stage that makes our scenario qualitatively different from
others.

We note that by itself, the possibility of an early epoch with
nearly Minkowski metric
is 
nothing new. Such an epoch is characteristic of a number 
of alternatives 
to  inflation, e.g., 
ekpyrotic models~\cite{Khoury:2001wf,Lehners:2008vx, Hinterbichler:2011qk} 
and Galilean Genesis~\cite{Creminelli:2010ba}. 
In these cases, however, perturbations get frozen 
once and for all at some time
before the hot era, and in this sense their evolution is 
similar to that in the inflationary theory. 
Not surprisingly, 
shapes of resulting bispectra are also 
similar to ones obtained in
versions of
inflationary scenario~\cite{Lehners:2007wc, Hinterbichler:2012mv}. 
In our case, however, we expect
qualitatively distinct shapes of non-Gaussianities.

An example of cosmology with the intermediate Minowskian evolution
of scalar perturbations is a version of conformal
rolling scenario. In that 
scenario~\cite{Hinterbichler:2011qk,Creminelli:2010ba, Rubakov:2009np}, 
one assumes that at early times,
the space-time geometry
is (effectively) Minkowskian, and the theory possesses
conformal symmetry. This symmetry is spontaneously broken
by a field $\rho$ of non-zero conformal weight $\Delta$, which homogeneously
rolls away
from the conformal point, $\rho_c \propto |\eta|^{-\Delta}$, 
where $\eta$ is the conformal time, $\eta<0$. One also assumes
that there is another field $\theta$ of zero conformal weight. By
conformal symmetry, the linear evolution of perturbations
$\delta \theta$ in the background $\rho_c (\eta)$ is the same as
that of massless scalar field in 
the de Sitter space-time. And the result is the same:
perturbations $\delta \theta$ get frozen at late rolling stage
and have flat power spectrum then.

At some point, conformal symmetry becomes irrelevant, and $\rho_c$ 
settles down to some condensate value. After that, the field $\theta$ evolves as
 massless scalar field minimally coupled to gravity. 
Further behavior of its perturbations depends on the 
cosmological evolution at that time. Generically, there are two
sub-scenarios. One is that
the modes of interest 
are already superhorizon in the conventional sense, and perturbations 
$\delta \theta$ remain frozen until the hot era. 
Non-linear effects in this sub-scenario have been considered in
Refs.~\cite{Hinterbichler:2012mv,Libanov:2010nk,Libanov:2010ci, Creminelli:2012qr}. 
The situation of our primary interest occurs in the second sub-scenario. 
It assumes that 
perturbations $\delta \theta$ evolve non-trivially during long enough epoch
between the end of conformal rolling and horizon exit~\cite{Libanov:2011hh}. 
The metric must be nearly Minkowskian at this stage, otherwise 
the flat power spectrum 
generated at conformal rolling
would be grossly modified. This second sub-scenario is realized, e.g.,
if conformal rolling occurs and ends up
early at the ekpyrotic contraction stage; the nearly Minkowskian evolution
of perturbations $\delta \theta$ takes place in that case from the end
of conformal rolling 
almost to the bounce. As deviations from the Minkowski metric 
become strong enough, perturbations $\delta \theta$ leave the horizon, 
pass unaffected through the bounce and remain unchanged until 
the beginning of the hot era. At some time at the radiation dominated 
stage, they convert into adiabatic fluctuations. 
We assume that the latter literally inherit properties 
of perturbations $\delta \theta$.


Hereafter, we prefer not stick to any particular
early Universe model and 
consider the general evolutionary picture of Fig.~\ref{evolution}. 
In addition, we assume 
that perturbations $\delta \theta$ are massless and non-interacting 
during the intermediate Minkowskian stage. 
In this situation, they are related in a simple way 
to the scale-invariant amplitude $\delta \theta ({\bf k}, \eta_*)$ existing at 
the beginning of the intermediate stage $\eta_*$, 
namely\footnote{One may wonder
 if this 
simple form of the solution survives the horizon crossing at 
$\eta \sim \eta_{ex}$, when
deviation from Minkowski metric becomes strong.
The fact that this is indeed so, at least 
in some 
cosmological scenarios, has been demonstrated in Appendix~A of 
Ref.~\cite{Libanov:2011hh}. There, the 
free propagation of perturbations $\delta \theta$ was considered
in the Universe filled with matter with the 
super-stiff equation of state, $p =w\rho$, $w\gg 1$, like 
in ekpyrotic models. It was shown 
that in the limit $w\rightarrow \infty$, the 
evolution of perturbations $\delta \theta$ is effectively Minkowskian 
all the way 
up to times $\eta \rightarrow 0$, when perturbations $\delta \theta$ are 
in the superhorizon regime.},
\begin{equation}
\nonumber 
\delta \theta ({\bf k}, \eta) =\delta \theta ({\bf k}, \eta_*)\cos k(\eta_* -\eta) \; .
\end{equation}
Here ${\bf k}$ is the conformal
momentum of a scalar mode,
$\eta$ is the conformal time running from $\eta_*$ 
to the horizon exit $\eta_{ex}$. Without loss 
of generality we set $\eta_{ex}=0$, then after the horizon exit
one has 
\[
\delta \theta ({\bf k}) =\delta \theta ({\bf k}, \eta_*)\cos k\eta_* \; .
\]
Due to this relation, the form of the primordial bispectrum 
is related in a simple way to
the initial shape function $A(k_1,k_2,k_3)$ generated before 
the Minkowskian evolution. It reads in terms 
of the primordial Newtonian potential $\Phi ({\bf k})$ 
\begin{equation} 
\label{bimoment}
\langle \Phi ({\bf k}_1) \Phi ({\bf k}_2) \Phi ({\bf k}_3) \rangle =
\left( \frac{\pi {\cal P}_{\Phi}}{2}\right)^{3/2} A(k_1,k_2,k_3) 
\delta \left(\sum_i {\bf k}_i \right) \cos (k_1\eta_*) \cos (k_2 \eta_*) 
\cos (k_3 \eta_*) \; .
\end{equation} 
Here ${\cal P}_{\Phi}$ is the power spectrum of
the scalar perturbation $\Phi ({\bf k})$, and we neglect the 
spectral tilt in what follows; delta function stands for momentum conservation; 
the normalization factor is chosen for future convenience. 
Until Section 3 we do not concretize the form of the shape 
function $A(k_1,k_2,k_3)$, considering it as an 
arbitrary slowly varying function of momenta. 
This is to show that effects due to the Minkowskian evolution 
are fairly generic. 

The modification of the bispectrum due to the intermediate Minkowskian
stage shows up
in the Cosmic Microwave Background (CMB) temperature 
fluctuations. One effect is the suppression of the CMB bispectrum 
by the inverse duration squared of 
this stage. 
Accordingly, 
the bispectrum vanishes as the duration of this stage tends to infinity. 
On the other hand, the bispectrum 
can be of the observable size for sufficiently short Minkowskian evolution. 
There is another interesting manifestation: the shape of the 3-point function 
undergoes oscillations in the multipole $l$ space. 
These are characterized by 
nearly constant amplitude in the range
$l_2 \leq l_3 \leq l_1+l_2$ (with the convention
$l_1 \leq l_2 \leq l_3$; the bispectrum vanishes for $l_3 > l_1+l_2$), 
unlike in inflationary scenarios with 
non-Bunch--Davies initial conditions~\cite{Holman:2007na, Chialva:2011hc,Meerburg:2009ys, Agullo:2010ws, Chen:2010bka}. 
This makes us argue that known constraints 
derived from the Planck data 
are not directly relevant to the case we study in the present paper.

This paper is organized as follows. In Section~2 
we study the imprint of the 
bispectrum~\eqref{bimoment} on CMB. 
We give analytic calculations of the CMB bispectrum for the 
cases of sufficiently long and relatively short intermediate 
Minkowskian stage. 
Details of calculations can be found in Appendices~A, B, C and D. In Section 3, we 
discuss the observability issues of the bispectrum with the Planck data. We summarize 
our findings in Section 4.

\section{Imprint on CMB}

\subsection{Generalities}

In the multipole representation, CMB temperature coefficients are defined by
\begin{equation}
\nonumber
a_{lm} =\int d {\bf n} \delta T({\bf n}) Y^{*}_{lm} ({\bf n})\; ,
\end{equation}
where $\delta T({\bf n})$ is the temperature fluctuation in the direction
 ${\bf n}$ in the sky, and $Y_{lm} 
({\bf n})$ are spherical harmonics. 
Coefficients $a_{lm}$ are related to the primordial Newtonian potential $\Phi({\bf k})$ by 
\begin{equation}
\nonumber
a_{lm} =4\pi i^{l} \int \frac{d {\bf k}}{(2\pi)^{3/2}} \Delta_l (k\eta_0) \Phi ({\bf k}) Y^{*}_{lm} (\hat{{\bf k}}) \; ,
\end{equation}
where $\Delta_l (k\eta_0)$ are standard CMB transfer functions, and $\eta_0$ is the present horizon radius; $\hat{{\bf k}}$ is the direction of the 
momentum ${\bf k}$. 
Recall that $\Delta_l (y) \propto j_l(y)$, where $j_l$ is the spherical Bessel
function. 
Now, using Eq.~\eqref{bimoment} and performing 
the integration over the third momentum ${\bf k}_3$, we obtain for the CMB bispectrum 
\begin{equation}
\label{bigeneral}
\begin{split}
\langle a_{l_1m_1} a_{l_2m_2} a_{l_3m_3} \rangle &
=i^{l_1+l_2+l_3} {\cal P}_{\Phi}^{3/2}\int d {\bf y}_1 d{\bf y}_2\Delta_{l_1} (y_1) 
\Delta_{l_2} (y_2) \Delta_{l_3} (|{\bf y}_1+{\bf y}_2|) \times \\
& \times Y^{*}_{l_1m_1} (\theta_1, \phi_1) Y^{*}_{l_2 m_2} (\theta_2, \phi_2) Y^{*}_{l_3m_3} (\theta_3, \phi_3) \times \\
& \times \cos (y_1 z) \cos (y_2 z) \cos (|{\bf y}_1+{\bf y}_2|z) A(y_1,y_2,y_3) \; ,
\end{split}  
\end{equation}
where the angles $(\theta_1, \phi_1)$, $(\theta_2, \phi_2)$ and 
$(\theta_3, \phi_3)$ correspond to directions of vectors
${\bf y_1}$, ${\bf y_2}$ and ${\bf y_3} = -({\bf y_1} + {\bf y_2})$, 
respectively. Here we introduced notations 
\begin{equation}
\nonumber
{\bf y}_i \equiv {\bf k}_i \eta_0, \qquad z \equiv -\frac{\eta_*}{\eta_0} \; ,
\end{equation}
and 
\begin{equation}
\nonumber
A(y_1, y_2, y_3) \equiv \frac{A(k_1,k_2,k_3)}{\eta^6_0} \; .
\end{equation}
For the sake of concreteness, we assume the ordering $l_1 \le l_2\le l_3$ in what follows.
And one more comment is in order, before we dig into details of calculations. 
The parameter $z$ is naturally large, since we are interested in the long Minkowskian evolution, meaning that 
$z \gg 1$. Leaving aside fine-tuning issues, however, 
the parameter $z$ is allowed to be not exceedingly large,
so we will also consider the case 
$z \gtrsim 1$. Going into even 
smaller values of $z$ is most likely not legitimate, as 
one may ruin the prediction for 
the scale-invariant power spectrum\footnote{In more detail, 
due to the presence of the intermediate stage, the 
power spectrum gets modified as ${\cal P}_{\Phi} \propto \cos^2 yz$. 
On the observational side this implies 
order $1/(lz)$ correction to the angular power spectrum $C_{l}$. For intermediate 
stage with the duration $z$ larger than unity, the correction is within the 
error bars due to the cosmic varaince $\sim 1/\sqrt{2l+1}$. For smaller $z$, however, this is no longer the case.}. Note that $y_iz \sim l_iz \gg 1$ for 
interesting multipoles in any case.

Despite the
complicated structure of the bispectrum, one can perform 
analytical integration over angles $\theta_i$ and $\phi_i$. 
To this end, we write the product of three 
rapidly oscillating cosines as follows
\begin{equation}
\label{cosdistort}
\begin{split}
\cos (y_1 z) \cos (y_2 z) \cos (y_3 z) &=\frac{1}{4} 
[ \cos \{(y_1 +y_2 +y_3) z\} +\cos \{(y_1 +y_2 -y_3) z\} +\\ 
&+\cos \{(y_1 -y_2 -y_3) z \} +\cos \{(y_2-y_1-y_3) z\}] \; ,
\end{split}
\end{equation}
where $y_3 = |{\bf y}_1 + {\bf y}_2|$. 
The contribution due to the first cosine in square brackets 
is always negligibly small. The reason is that 
this term rapidly oscillates 
for all relevant values of the parameter $z$ and CMB multipole numbers. 
Similar story happens 
with the contributions due to the other three terms, 
except for small regions where oscillations are damped. 
For concreteness, we pick the second cosine in 
the r.h.s. of Eq.~\eqref{cosdistort}. 
Its argument has an extremum at $y_3=y_1 + y_2$, i.e., for ${\bf y}_1$
parallel to ${\bf y}_2$. Expanding the argument up to the 
second order in  
$\delta \theta_1 \equiv \theta_1 -\theta_2$ and $\delta \phi_1 \equiv \phi_1 -\phi_2$, we obtain
\begin{equation}
\label{argcosine}
y_1 +y_2 -|{\bf y}_1 +{\bf y}_2| \approx \frac{y_1 y_2}{2(y_1 +y_2)} 
\{\delta \theta^2_1 +\sin^2 \theta_2 \delta \phi^2_1 \} \; .
\end{equation}
One observes that oscillations are damped in the region 
\begin{equation}
\label{region}
|\delta \theta_1| \lesssim \frac{1}{\sqrt{N}} \quad \mbox{and} 
\quad |\delta \phi_1| \lesssim \frac{1}{\sin \theta_2 \sqrt N} \; ,
\end{equation}
where $N$ is the large dimensionless parameter 
\begin{equation}
\label{N}
N=\frac{y_1 y_2 z}{2(y_1+y_2)} \; .
\end{equation} 
Thus, for 
calculating the integral~\eqref{bigeneral}, 
the saddle point technique is naturally employed. 
Depending on the duration of the intermediate stage, we distinguish 
two cases. For $z \gg l$, 
one keeps only the product of three rapidly oscillating cosines in 
Eq.~\eqref{bigeneral} when 
finding saddle points. For shorter duration of the intermediate stage, 
we should account for variations of spherical harmonics as well. 

Conservatively, the 
two cases are separated at $z \sim l_{3}$, 
where $l_3$ is the largest multipole number, according to our conventions. 
In practice, the calculation neglecting the variation of the spherical
harmonics can be extended
down to $z \sim l_{1}$ (recall that $l_{1} =\mbox{min}(l_1,l_2,l_3)$). 
Let us argue for this. We first focus on the contribution due to the second term in Eq.~\eqref{cosdistort}. 
We are free to integrate first over angles $\theta_1$ and $\phi_1$ 
associated with the smallest multipole number $l_1$. 
In that case, the spherical harmonic 
$Y_{l_2m_2}$ is not in the game. Variation of the spherical harmonic $Y_{l_1m_1}$ for sufficiently small values 
of $\delta \theta_1$ is estimated as 
$|\delta Y_{l_1m_1}| \sim |l_1 Y_{l_1m_1} \delta \theta_1|$. Hence, the
inequality 
$|\delta Y_{l_1m_1}| \ll |Y_{l_1m_1}|$ 
holds in the region~\eqref{region}, provided that 
$z$ is greater than  $l_1$.
The estimate for the
variation of the spherical harmonic with the largest multipole number 
reads $|\delta Y_{l_3m_3}| \sim |l_3 Y_{l_3m_3} \delta \theta_3| \sim |l_1 Y_{l_3m_3} \delta \theta_1|$. 
The second estimate here follows from the triangle relation ${\bf k}_1+{\bf k}_2+{\bf k}_3=0$. 
See also Eqs.~\eqref{theta3} and~\eqref{phi3} of Appendix~A. We conclude 
that the variation of the spherical harmonic $Y_{l_3m_3}$ is 
also sufficiently small 
in the region~\eqref{region} for $z$ larger than $l_1$. 

This discussion is extrapolated to the 
contribution of the fourth cosine in Eq.~\eqref{cosdistort} in a straightforward manner. 
On the other hand, above argument does 
not work for the contribution of the third term there. 
However, it gives 
contribution negligible as compared to the other two in the case of the 
hierarchy between multipole numbers, $l_1 \ll l_3$. 
The reason is that the corresponding integral is 
saturated in the region $y_1 \approx y_2 + |{\bf y}_1 + {\bf y_2}|\equiv y_2+y_3$ 
(i.e., $y_2 < y_1$ and vector ${\bf y}_2$ anti-parallel to ${\bf y}_1$). 
The transfer functions $\Delta_{l_2} (y_2) \propto j_{l_2}(y_2)$ and
 $\Delta_{l_3} (y_3) \propto j_{l_3}(y_3)$ do not vanish only at 
$y_2 \geq l_2+1/2$ and $y_3 \geq l_3+1/2$, respectively. Both 
the shape function $A(y_1,y_2,y_3)$ and transfer function 
$\Delta_{l_1} (y_1)$ are suppressed for $y_1\approx y_2+y_3> l_3 \gg l_1$. 
Also, the product of the three transfer functions is a rapidly oscillating function 
of variables $y_i$. 
These effects lead to quite strong suppression of the term we discuss, 
see Appendix~B for details. 

We consider the case of long intermediate stage, i.e. $z \gg l_1$, 
in the following Subsection. We generalize the results to the case of 
arbitrary $z$ (but still $z \gtrsim 1$) in Subsection 2.3.

\subsection{Saddle-point calculation: regime $z \gg l_1$}

We first consider the case of sufficiently long intermediate stage,
$z \gg l_1$. In this situation, we neglect 
variations of spherical harmonics and find the saddle points directly
from Eq.~\eqref{argcosine}. We focus 
on the contribution due to the second cosine in the r.h.s. 
of Eq.~\eqref{cosdistort}; contributions due to the 
third and fourth cosines are then obtained  by permuting 
multipole numbers $(l_1,m_1)$ and $(l_2, m_2)$ with $(l_3, m_3)$. 
The saddle point is at 
$\delta {\theta}_1 =0$ and $\delta { \phi}_1=0$ in terms of variables 
$\delta \theta_1=\theta_1-\theta_2$ and $\delta \phi_1=\phi_1-\phi_2$.
When integrating in the vicinity of the saddle point, 
we expand the bispectrum~\eqref{bigeneral} in a series in
$1/z$.

Naively, the first term  in this expansion is of the order 
$1/z =-\eta_0/\eta_* $. However, the suppression is actually stronger. 
To show this, we consider the inner integral over 
angles $\theta_1$ and $\phi_1$ in Eq.~\eqref{bigeneral}, 
write the product of three cosines as in 
Eq.~\eqref{cosdistort} and pick the second cosine in the r.h.s. there. 
To the leading order in $1/z$ we neglect the dependence on
$\delta \theta_1$, $\delta \phi_1$ of all other factors in the integrand in 
Eq.~\eqref{bigeneral}. Using Eq.~\eqref{argcosine}, we obtain 
\begin{equation}
\nonumber
\begin{split}
& \int d \theta_1 d\phi_1  \cos \Bigl[\frac{y_1y_2 z}{2(y_1 +y_2)} 
\Bigl (\delta \theta^2_1 +\sin^2 \theta_2 
\delta \phi^2_1\Bigr) \Bigr]=\\ &= \frac{2(y_1+y_2)}{y_1 y_2 z \sin \theta_2}\Bigl[\Bigl( \int^{+\infty}_{-\infty} 
d u \cos u^2 \Bigr)^2 -\Bigl(\int^{+\infty}_{-\infty} 
du \sin u^2 \Bigr)^2 \Bigr] = 0\; .
\end{split}
\end{equation}
The last equality follows from the relation
\begin{equation}
\nonumber 
\int^{+\infty}_{-\infty} \cos u^2 du = \int^{+\infty}_{-\infty} \sin u^2 du \; .
\end{equation}

We see that
the leading contribution to the bispectrum is of the order $1/z^2$. To 
calculate the 3-point 
function to this order, one expands all 
slowly varying functions of $\theta_1$, $\phi_1$ in Eq.~\eqref{bigeneral}
up to terms quadratic in $\delta \theta_1$ and $\delta \phi_1$. 
At the same time, the argument of the rapidly oscillating 
cosine must be expanded up to orders $N (\delta \theta_1)^4$, $N(\delta \phi_1)^4$ and  
$N (\delta \theta_1)^2 (\delta \phi_1)^2$. See Appendix~A for 
explicit formulae. 
It is convenient to split the integral in Eq.~\eqref{bigeneral} 
into the sum of two contributions, 
\begin{equation}
\label{formality}
\langle a_{l_1 m_1} a_{l_2 m_2} a_{ m_3} \rangle =\langle a_{l_1 m_1} a_{l_2 m_2} a_{l_3 m_3} \rangle_0 
+\delta \langle a_{l_1 m_1} a_{l_2 m_2} a_{l_3 m_3} \rangle \; .
\end{equation} 
The first term in the r.h.s. involves 
variations of all functions, except for $A(y_1, y_2,y_3)$ 
and $\Delta_{l_3} (y_3)$. 
By writing it with the subscript $0$, we anticipate
 that it gives the dominant contribution to the 
3-point function. The second term in the r.h.s. accounts for variations of 
functions $A(y_1,y_2,y_3)$ and $\Delta_{l_3}(y_3)$. 
The expression for the main contribution to the order $1/z^2$ is given by
\begin{equation}
\label{bileading}
\begin{split}
\langle a_{l_1m_1} a_{l_2m_2} a_{l_3m_3} \rangle_0 &=i^{l_1+l_2+l_3} 
\cdot \frac{\pi {\cal P}^{3/2}_{\Phi}}{4z^2} 
\cdot \int dy_1 dy_2 \Delta_{l_1} (y_1) \Delta_{l_2} (y_2) \Delta_{l_3} (y_1+y_2)\times \\ 
& \times A(y_1,y_2,y_1+y_2) \cdot B^{l_1m_1}_{l_2m_2;l_3m_3} \cdot 
\Bigl \{ y^2_2l_1(l_1+1) +y^2_1l_2(l_2+1)  -\\ 
&-2y_1y_2 \sqrt{l_1(l_1+1)} \sqrt{l_2(l_2+1)} 
B^{l_1,1}_{l_2,-1;l_3,0} \Bigl(B^{l_10}_{l_20;l_30} \Bigr)^{-1} +2y_1y_2\Bigr \}+\\ 
&+ (l_1,m_1\leftrightarrow l_3,m_3) + (l_2,m_2 \leftrightarrow l_3,m_3) \; ;
\end{split}  
\end{equation}
we refer the reader to Appendix~A for details of calculations. 
Constants $B^{l_1 m_1}_{l_2 m_2; l_3 m_3}$ entering the integrand 
in Eq.~\eqref{bileading} are given by
\begin{equation}
\label{B1}
B^{l_1 m_1}_{l_2 m_2; l_3 m_3} =\int d \phi d \theta \sin \theta~  Y^{*}_{l_1 m_1} (\theta, \phi) 
Y^{*}_{l_2 m_2} (\theta, \phi) Y^{*}_{l_3 m_3} (\pi-\theta, \pi +\phi) \; .
\end{equation}
They can be expressed via the Wigner $3j$-symbols,
\begin{equation}
\label{B}
B^{l_1 m_1}_{l_2 m_2; l_3 m_3}=
(-1)^{l_3}\sqrt{\frac{(2l_1+1)(2l_2+1)(2l_3+1)}{4\pi}}\left(
\begin{array}{ccc} 
l_1 & l_2 & l_3\\
0 & 0 & 0
\end{array}
\right) 
\left(
\begin{array}{ccc} 
l_1 & l_2 & l_3\\
m_1 & m_2 & m_3
\end{array}
\right) \; .
\end{equation}
Note that the term in the bispectrum written explicitly in
Eq.~\eqref{bileading} 
is symmetric 
under the interchange of multipole numbers 
$(l_1,m_1)$ and $(l_2, m_2)$, as it should. 
Furthermore, its form exhibits statistical isotropy, 
since this is the 
property of coefficients $B^{l_1m_1}_{l_2m_2;l_3m_3}$; see Appendix~A for 
discussion of this point. These two 
observations
serve as cross-checks of our computations.

As we pointed out in the end of Subsection 2.1, Eq.~\eqref{bileading}
can be used down to $z \sim l_1$, even though the saddle point calculation
of the term with permutation $(l_1,m_1) \leftrightarrow (l_3,m_3)$
is not justified for $z \sim l_1$ and $l_1 \ll l_3$. In fact, the argument in 
the end of Subsection 2.1 suggests that this term (the first term in the
last line of Eq.~\eqref{bileading}) is suppressed at $l_1 \ll l_3$. 
We check this explicitly in Appendix B.


The correction to the bispectrum due to variations 
of shape and transfer functions $A(y_1,y_2,y_3)$ 
and $\Delta_{l_3}(y_3)$ is given by
\begin{equation}
\label{bicorrection}
\begin{split}
\delta \langle a_{l_1 m_1} a_{l_2 m_2} a_{l_3 m_3} \rangle & = i^{l_1+l_2+l_3} \cdot \frac{\pi {\cal P}^{3/2}_{\Phi}}{2z^2}
\cdot  \int dy_1 dy_2 
\Delta_{l_1} (y_1)  \Delta_{l_2} (y_2) \Delta_{l_3} (y_1+y_2) \times \\ & 
\times A (y_1, y_2, y_1+y_2) \cdot B^{l_1 m_1}_{l_2 m_2; l_3 m_3} \times \\
& \times y_1y_2 \cdot \Bigl(\frac{\partial 
\ln A (y_1, y_2, y_3)}{\partial \ln y_3} +\frac{d \ln \Delta_{l_3} (y_3)}{d \ln y_3} 
 \Bigr)_{y_3=y_1+y_2} \\
& + (l_1,m_1\leftrightarrow l_3,m_3) + (l_2,m_2 \leftrightarrow l_3,m_3)\; .
\end{split} 
\end{equation} 
Since $A(y_1,y_2,y_3)$ is a slowly varying function of its arguments, we estimate the corresponding term as 
$\partial \ln A/ \partial \ln y_3 \sim 1$. The estimate for the second term 
in parenthesis of Eq.~\eqref{bicorrection} reads 
\begin{equation}
\nonumber 
\frac{d \ln \Delta_{l_3} (y_3)}{d \ln y_3} \sim y_3 
\cdot \frac{|\Delta_{l_3-1} (y_3) -\Delta_{l_3+1} (y_3)|}{|\Delta_{l_3} (y_3)|} \lesssim y_3 \; ,  
\end{equation}
which follows from the fact that rapidly varying part of
$\Delta_l (y)$ is proportional to the spherical Bessel function
$j_l (y)$ and from 
the relation for the derivative of the spherical Bessel function. 
Using crude estimate $y_i \sim l_i$, 
we conclude that the contribution~\eqref{bicorrection} is suppressed as compared to one in Eq.~\eqref{bileading} 
by at least a factor of $l^{-1}_{1}$. 

Equations~\eqref{bileading} and \eqref{bicorrection} give the explicit expression for the CMB 
3-point function to the order $1/z^2$.

\subsection{Saddle-point calculation: generic case}

If the duration of the intermediate stage is not particularly large, spherical harmonics can no longer be treated as smooth 
functions. Consequently, Eq.~\eqref{bileading} is not valid in this regime. 
Still, we employ the saddle point technique with the modification that
we now search for  
saddle points of the product of the rapidly 
oscillating cosines 
and spherical harmonics. To simplify the problem, we use the 
approximate form of spherical harmonics valid for large 
multipole numbers $l \gg 1$,  
\begin{equation}
\label{spherap}
Y_{lm} (\theta, \phi) =\frac{1}{\pi \sqrt{\sin \theta}} \cos \left[\left(l+\frac{1}{2}\right)\theta -\frac{\pi}{4} +\frac{\pi m}{2}
\right]e^{im\phi} +{\cal O} 
\left(\frac{1}{l} \right) \; .
\end{equation}
Clearly, this approximation enables one to perform
calculations to the leading order in multipole numbers 
$l_i$ only. Sketch  of calculations, tedious but fairly straightforward, 
is given in Appendix~C. Here we point out the main technical 
aspect: due to variations of spherical harmonics, 
saddle points in terms of the variable 
$\delta \theta_1 \equiv \theta_1 -\theta_2$ 
get shifted to 
$\delta { \theta}_1 \sim \frac{1}{z}$. 
This is consistent with the previous Subsection: we have for 
the saddle point value that $\delta \theta_1 \to 0$ as $z \to \infty$. 
The final expression for the CMB bispectrum reads
\begin{equation}
\label{bishort}
\begin{split}
\langle a_{l_1m_1} a_{l_2m_2} a_{l_3m_3} \rangle &=i^{l_1+l_2+l_3} \cdot \frac{\pi{\cal P}^{3/2}_{\Phi}}{4}  \cdot
B^{l_1m_1}_{l_2m_2;l_3m_3} \times \\ 
& \times \int  dy_1 dy_2 \cdot y^2_1 \cdot y^2_2 \cdot \Delta_{l_1} (y_1) \Delta_{l_2} (y_2) \Delta_{l_3} (y_1+y_2) 
 \cdot A(y_1,y_2,y_1+y_2) \times \\
& \times \frac{1}{N} \cdot \sin \Bigl \{\frac{1}{2} \cdot \frac{y^2_2l^2_1 +y^2_1l^2_2}{y_1y_2(y_1+y_2) z} \Bigr \} \cdot \cos 
\Bigl \{ \frac{l_1l_2}{(y_1+y_2)z} \Bigr \} \times \\
& \times \Bigl[1-
\cot \Bigl \{ \frac{1}{2} \cdot \frac{y^2_2l^2_1 +y^2_1l^2_2}{y_1y_2(y_1+y_2) z}   \Bigr \} \cdot 
\tan \Bigl \{ \frac{l_1l_2}{(y_1+y_2)z} \Bigr \} \cdot \frac{B^{l_1,1}_{l_2,-1;l_30}}{B^{l_10}_{l_20;l_30}}\Bigr]\\
&  + (l_1,m_1\leftrightarrow l_3,m_3) + (l_2,m_2 \leftrightarrow l_3,m_3)\; ,
\end{split}
\end{equation}
where the large parameter $N$ is given by Eq.~\eqref{N}. 
Let us comment on this formula. First, in the limit 
$z \gg l_{1}$ we get back to the expression~\eqref{bileading}, as we 
should. This is a simple cross-check of our computation. 
Second, in the case 
$1\ll z \ll l_1$ the suppression factor is $1/N \sim 1/(l_{1} z)$ 
as compared to $1/z^2$ in Eq.~\eqref{bileading}. 
This implies stronger squeezening of the bispectrum for relatively small $z$.

\subsection{Flattened triangle limit and away from it}

The 3-point function given by Eq.~\eqref{bileading} or Eq.~\eqref{bishort} 
has several peculiar features. 
Our first observation concerns the behaviour of the bispectrum in 
the flattened triangle
regime where $l_1 +l_2 - l_3\ll l_1$ (recall our convention
$l_1 \leq l_2 \leq l_3$). 
For the sake of concreteness, we specify to the case of long intermediate stage 
studied in Subsection 2.2. In the large $l$ approximation, 
the expression in curly brackets 
in Eq.~\eqref{bileading} takes the form
\begin{equation}
\label{brackets}
 l^2_1 \cdot l^2_2 \cdot 
\Bigl(\frac{y_1}{l_1} -\frac{y_2}{l_2} \Bigr)^2  +2y_1y_2 \cdot l_1l_2 \cdot  \Bigl[
1- B^{l_1,1}_{l_2,-1;l_3,0} (B^{l_10}_{l_20;l_30})^{-1} \Bigr]  \; .
\end{equation} 
Borrowing results from the next Section, we note that the first term here 
is suppressed due to the 
effective relation ${y_1}/{l_1} ={y_2}/{l_2}$ 
which holds in the flattened triangle regime modulo 
corrections of the order $1/\sqrt{l_{1}}$. Partial cancellation occurs also in the second term. It is 
due to the relation
\begin{equation}
\nonumber 
B^{l_1,1}_{l_2,-1;l_30}=\frac{\sqrt{l_1(l_1+1)}\sqrt{l_2(l_2+1)}}{(l_1+1)(l_2+1)} B^{l_10}_{l_20;l_30} \; ,
\end{equation}
which follows from the analogous relation between the
$3j$-symbols. Thus,
our bispectrum is suppressed by the factor $\sim 1/l_{1}$, as
compared to naive estimate, for configurations with 
 $l_1 +l_2 - l_3\ll l_1$.

The bispectrum is suppressed also away from the flattened triangle limit, 
but for different reason. 
The source of this suppression is the integral over three transfer functions, i.e.   
\begin{equation}
\nonumber
\int d y_1 d y_2 \Delta_{l_1} (y_1) \Delta_{l_2} (y_2) \Delta_{l_3} (y_1+y_2)... \; .
\end{equation}
We will perform the detailed analysis of this integral in Section~3. 
Here let us make a qualitative observation. 
The integral over variables $y_i$ is saturated roughly 
at $y_i \sim l_i$. Somewhat loosely, we set arguments of transfer functions at $l_1$, $l_2$ and $l_1+l_2$. 
Then the bispectrum 
is roughly proportional to
the product $\Delta_{l_1} (l_1) \Delta_{l_2} (l_2) \Delta_{l_3} (l_1+l_2)$.
The last factor here, $\Delta_{l_3} (l_1 + l_2) \propto j_{l_3}(l_1+l_2)$, 
vanishes for $l_1+l_2 < l_3$ and
undergoes rapid oscillations as function of $(l_1 + l_2- l_3)$ 
with the amplitude decaying away from the 
flattened triangle limit. Numerical 
estimates of Section 3 roughly agree with these expectations. Indeed, we will 
observe rapid oscillations naturally traced back to ones in 
the transfer functions. 
The qualification is that their 
amplitude is nearly constant 
in the interval $l_2\leq l_3 \leq l_1+l_2$, 
rather than decreasing. This is due to the cancellation in the flattened triangle limit 
we discussed above.

\section{Estimates and observational consequences}

So far we have
found indications for several potentially interesting features: oscillatory 
behaviour of the bispectrum, suppression both in the flattened 
triangle limit and away from it. 
The purpose of the present Section is to demonstrate the interplay 
between these effects.
Another purpose is to find the range 
of the parameter $z$ 
where the bispectrum is of 
observable size. To this end, 
we specify to the standard local form of the bispectrum generated 
before the long Minkowskian evolution,
\begin{equation}
\label{local}
A(k_1,k_2,k_3) = C \frac{\sum_i k^3_i}{\Pi_i k^3_i}\; , \;\;\;\;\;\;\;\;
A(y_1,y_2,y_3) = C \frac{\sum_i y^3_i}{\Pi_i y^3_i} \; .
\end{equation}
Here the constant $C$ is some combination of parameters inherent in
 the theory operating before the 
intermediate stage. Note that the bispectrum of the form~\eqref{local} is a generic prediction 
of a number 
of early Universe scenarios including curvaton~\cite{Zaldarriaga:2003my} and
conformal Universe~\cite{Hinterbichler:2012mv} models\footnote{More precisely, in both cases the form 
of the shape function is $A(y_1, y_2, y_3) \propto \frac{y^3_t}{\Pi_i y^3_i} \left( \frac{8}{9} -
\frac{2 \sum_{i<j} y_i y_j}{y^2_t}-\frac{1}{3} (\gamma +{\cal N} )\frac{\sum_i 2y^3_i}{y^3_t}\right)$, where 
$y_t=y_1+y_2+y_3$,  
$\gamma =0.577$ is the Euler constant, and ${\cal N}$ is a logarithmic 
factor. The latter is naturally large. Therefore, 
the last term in parenthesis dominates, 
implying the approximate 
form of the shape function as in 
Eq.~\eqref{local}. }. In both cases, the constant $C$ is not necessarily 
small, and can be as large as 
$10$ or $100$. Even larger values, however, 
are at risk of strong coupling between perturbations before
the intermediate Minkowski stage.

Because of the
statistically isotropic form of Eqs.~\eqref{bileading} and~\eqref{bishort}, 
it is convenient to introduce reduced bispectrum $b_{l_1l_2l_3}$ defined by 
\begin{equation}
\nonumber
\langle a_{l_1m_1} a_{l_2m_2} a_{l_3m_3} \rangle =(-1)^{l_3} B^{l_1m_1}_{l_2m_2;l_3m_3} b_{l_1l_2l_3} \; .
\end{equation}
We also make the 
following simplifications. First, we ignore the correction in Eq.~\eqref{formality} and focus on the main 
contribution~\eqref{bileading}, where we neglect 
terms with permutations. The latter are indeed small, as shown 
in Appendix~B.
Somewhat loosely, we also set functions $g^{T}_l (y) =\Delta_l (y)/{j_l (y)}$ 
equal to a constant 
in the integrands in Eqs.~\eqref{bileading} and~\eqref{bishort}, namely,
\begin{equation}
\label{replace}
g^{T}_l (y) \rightarrow g^{T}_{SW} \approx 1/3 \; .
\end{equation}
Although 
this approximation works only for relatively small
multipole numbers from the Sachs-Wolfe (SW) plateau of the CMB 
angular power spectrum, 
it is sufficient for our illustrational purposes. The rest of computations 
can be found in Appendix D. Final results are presented for the cases of long 
and short intermediate stage in Subsections 3.1 and 3.2, respectively.
\begin{figure}[tb!]
\begin{center}
\includegraphics[width=0.44\columnwidth, angle=0]{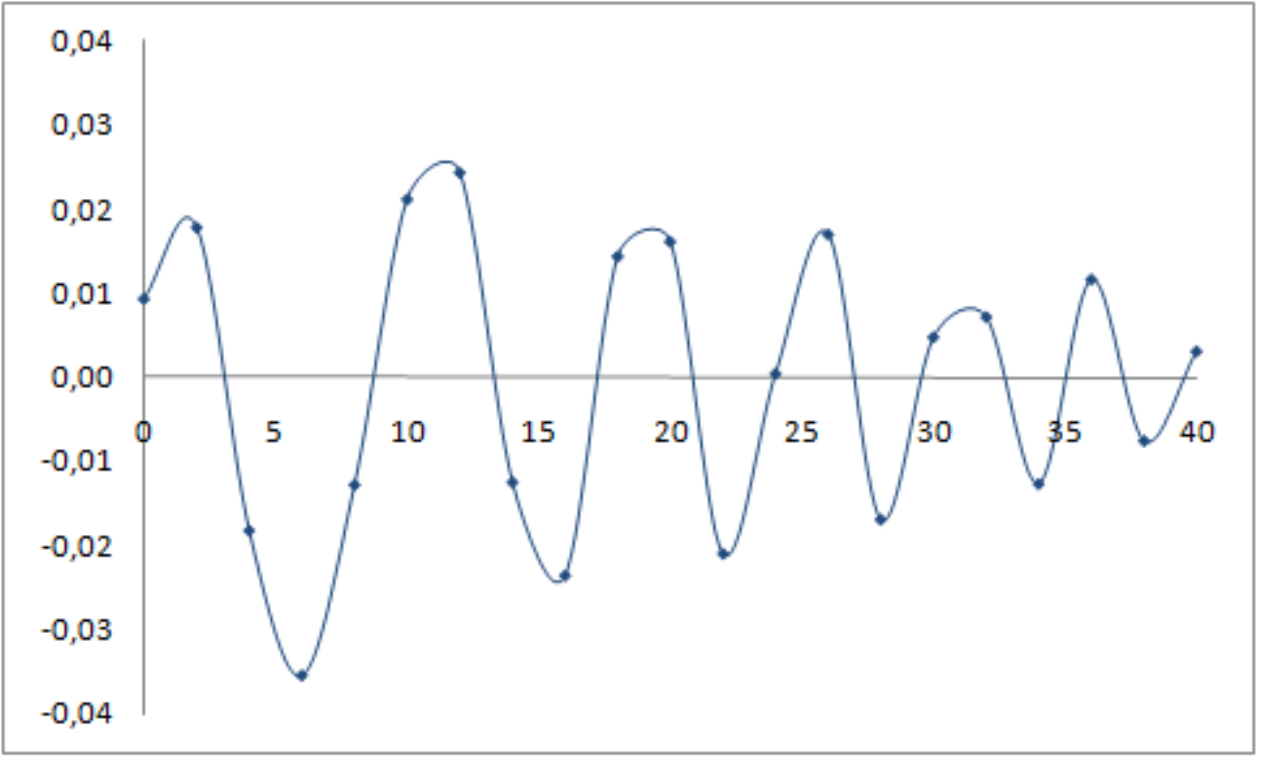}
\includegraphics[width=0.44\columnwidth, angle=0]{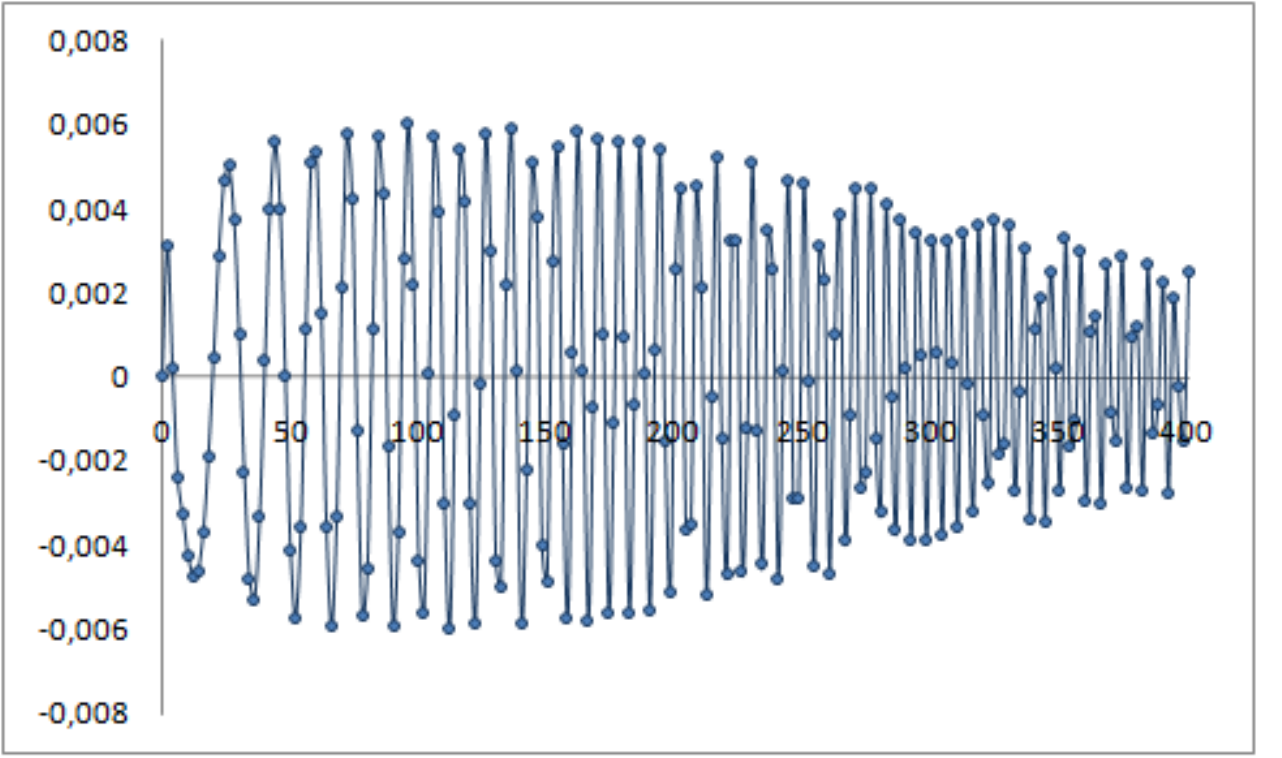}
\end{center}
\caption{Structure coefficients $c_{l_1l_2l_3}$ defined by Eq.~\eqref{clll} 
as  functions of $\Delta l \equiv l_1+l_2-l_3$ for  
 $l_2=600$, $l_1=50$ (left) and $l_1=l_2=600$ (right).}\label{c}.
\end{figure}
\begin{figure}[tb!]
\begin{center}
\includegraphics[width=0.44\columnwidth, angle=0]{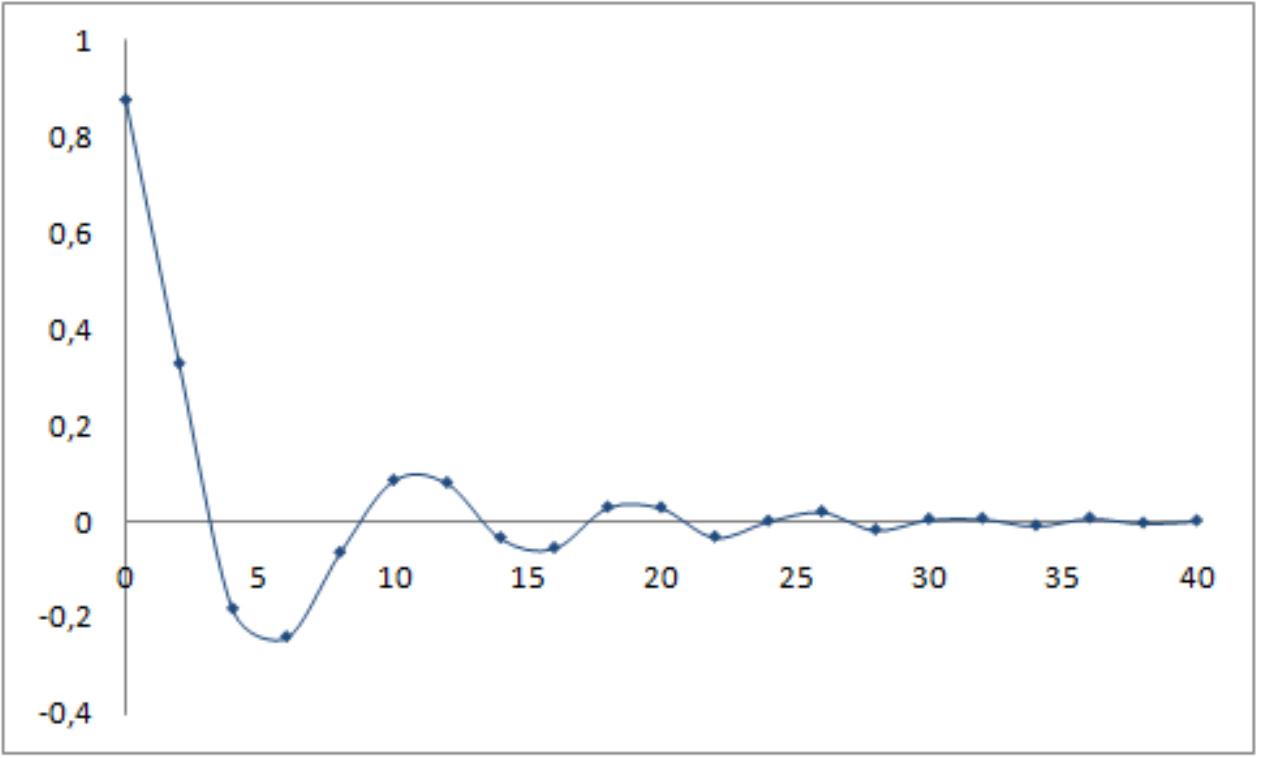}
\includegraphics[width=0.44\columnwidth, angle=0]{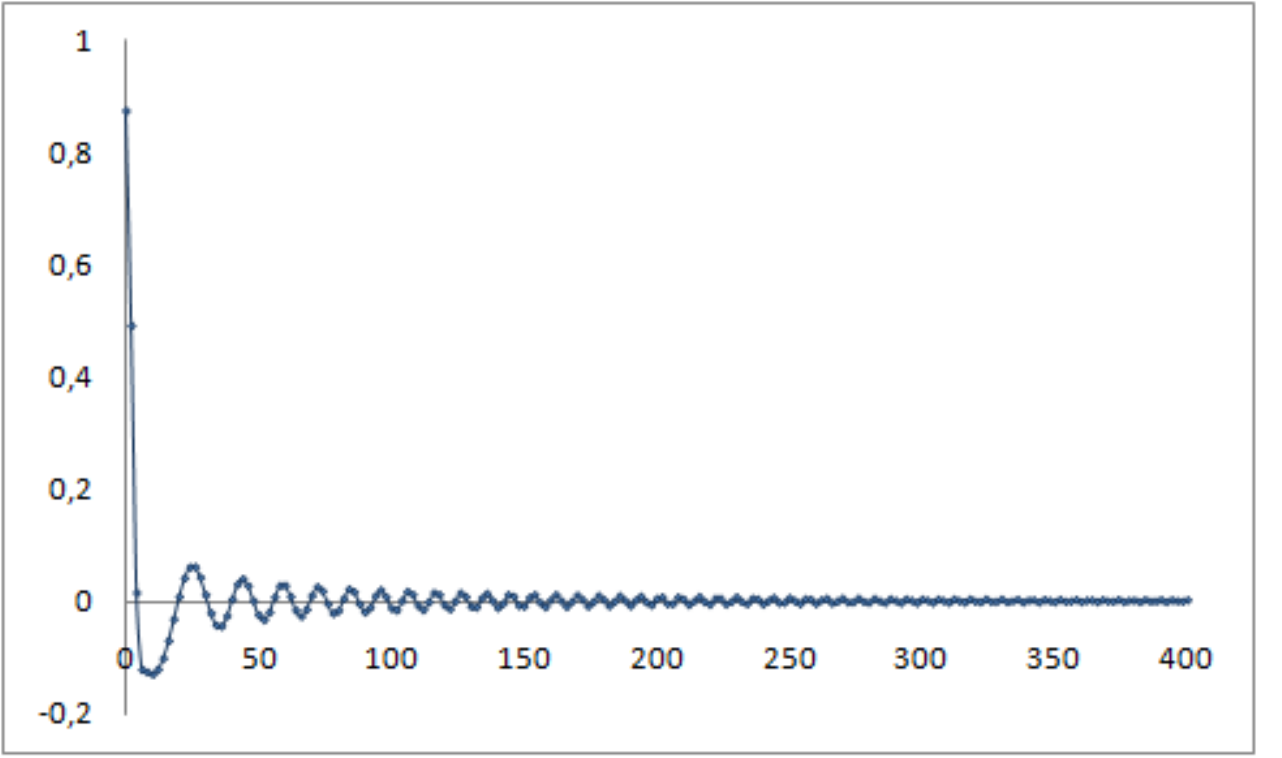}
\end{center}
\caption{Quantity ${\mathcal F}(l_1, \Delta l)$ defined by Eq.~\eqref{f} 
as function of the difference $\Delta l \equiv l_1+l_2-l_3$ 
for $l_1=50$ (left) and $l_1=600$ (right).}\label{fa}
\end{figure}
\begin{figure}[tb!]
\begin{center}
\includegraphics[width=0.44\columnwidth, angle=0]{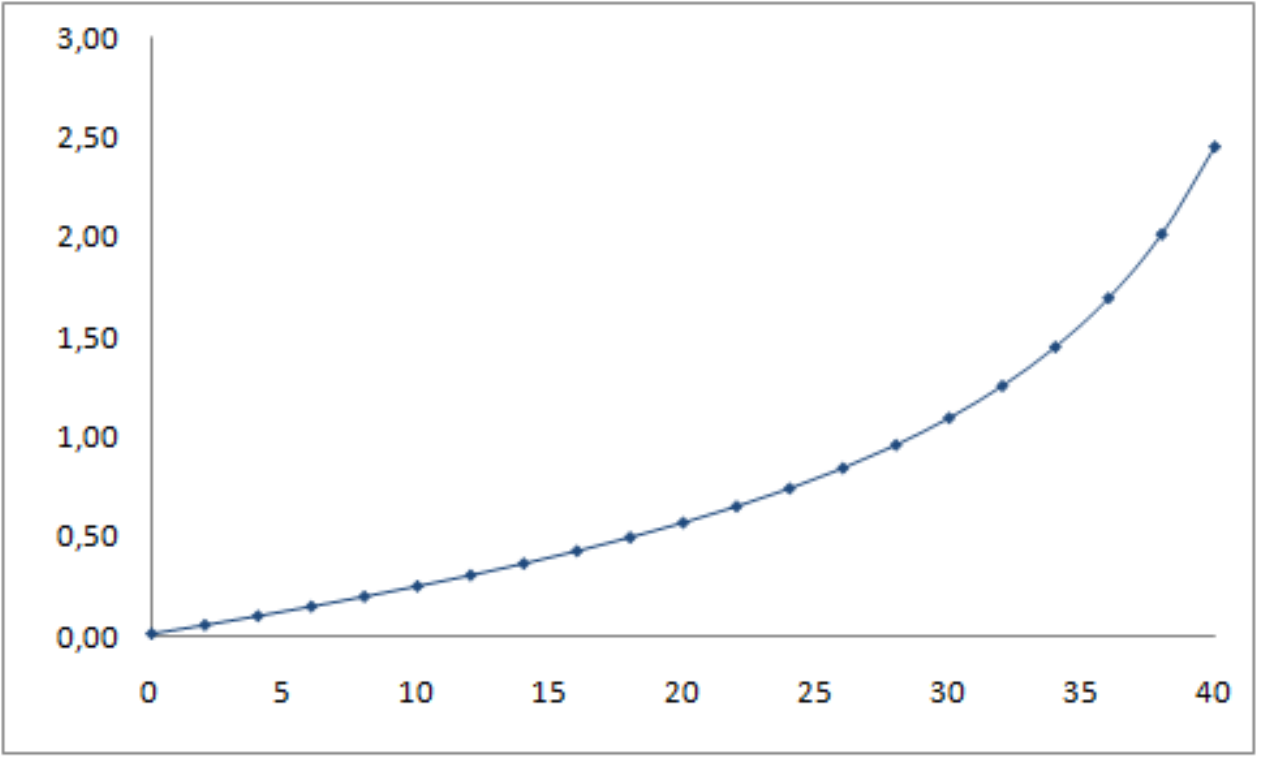}
\includegraphics[width=0.44\columnwidth, angle=0]{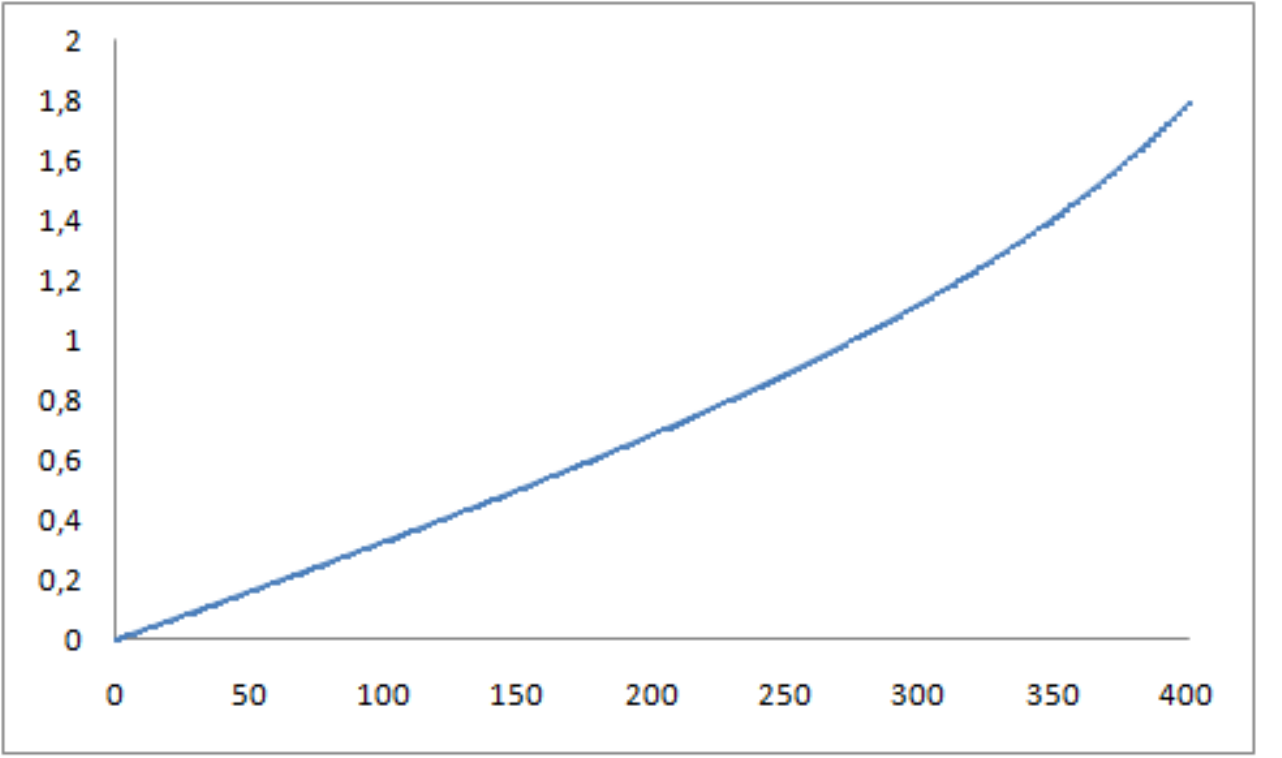}
\end{center}
\caption{Ratio $c_{l_1l_2l_3}/{\mathcal F}(l_1, \Delta l)$ as function 
of the difference $\Delta l \equiv l_1+l_2 -l_3$ 
for  $l_2=600$, $l_1=50$ (left) and $l_1=l_2=600$ (right).}\label{ratiocf}. 
\end{figure}

\subsection{Regime $z \gg l_{min}$}

First, we consider the case of long intermediate stage,
$z \gg l_1$. 
Using Eq.~\eqref{bileading} and following steps outlined in Appendix~D, we obtain 
the estimate for the reduced bispectrum
\begin{equation}
\label{mainestimate}
\begin{split}
|b_{l_1l_2l_3}| \sim  \frac{C}{30} \cdot \frac{{\cal P}^{3/2}_{\Phi}}{z^2}
\frac{|c_{l_1l_2l_3}|}{l^{5/2}_1 \cdot l^{3/2}_2 \sqrt{l_3}} \; ,
\end{split}
\end{equation}
where the quantities $c_{l_1l_2l_3}$ are negligible for
$l_3 > l_1+l_2$ and otherwise are given by
\begin{equation}
\label{clll}
c_{l_1l_2l_3}=  l_1\left[1+\frac{1}{2} 
\cdot \frac{\Delta l^2/l^2_1}{ 1-\Delta l/ l_1}-\frac{B^{l_30}_{l_1,1;l_2,-1}}{B^{l_3,0}_{l_1,0;l_2,0}} \right] 
{\mathcal F}(l_1, \Delta l) \; .
\end{equation}
Here $\Delta l \equiv l_1+l_2-l_3$ and 
the function ${\mathcal F}(l_1, \Delta l)$ 
involves the integral over the variable $u_1 \equiv \frac{y_1}{l_1+1/2}$, 
\begin{equation}
\label{f}
\begin{split}
 {\mathcal F}(l_1, \Delta l) &=\left(1-\frac{\Delta l}{l_1} \right)^{2}\int^{\infty}_1  \frac{du_1}{u^{9/2}_1 [u^2_1-1]^{1/4}  
\Bigl[u^2_1-\Bigl(1-\frac{\Delta l}{l_1} \Bigr)^2 \Bigr]^{1/4}} \times \\ &\times
 \Bigl[\cos \Bigl \{l_1 f(u_1) -(l_1 -\Delta l) f(\bar{u}_2) \Bigr \} 
-\sin \Bigl \{l_1 f(u_1)-(l_1-\Delta l) f(\bar{u}_2) \Bigr \} \Bigr] \; ,
\end{split}
\end{equation}
where $\bar{u}_2 =\frac{l_1}{l_3-l_2}u_1$. 
The non-trivial part of the bispectrum is encoded 
in the structure coefficients $c_{l_1l_2l_3}$. We plot them 
in Fig.~\ref{c} as functions 
of the difference $\Delta l$. For each plot, we keep 
smaller multipole numbers $l_1$ and $l_2$ fixed and vary the larger one $l_3$, 
and, consequently, $\Delta l=l_1+l_2-l_3$.  
The main property we observe are oscillations with roughly constant 
amplitude in the interval $l_2 \leq l_3 \leq l_1 + l_2$. 
To make things clearer, we also plot 
the function ${\mathcal F}(l_1, \Delta l)$ and the ratio $c_{l_1l_2l_3}/{\mathcal F}(l_1, \Delta l)$ in 
Figs.~\ref{fa} and~\ref{ratiocf}, respectively. 
Fig.~\ref{fa} demonstrates the tendency 
of the amplitude to decrease away from the flattened triangle limit, i.e. with  
growing  $\Delta l$, while Fig.~\ref{ratiocf} exhibits the opposite tendency. 
The two compensate each other and we get the nearly constant amplitude in the 
end. Another effect seen from Fig.~\ref{c} is the suppression
of the bispectrum  by a factor $l_1^{-1/2}$ 
as compared to the situation where the intermediate stage is absent. 
Indeed, the estimate for coefficients $c_{l_1l_2l_3}$ is 
roughly $c_{l_1l_2l_3} \sim 1$. Hence, the bispectrum~\eqref{mainestimate} scales as 
$b_{l_1l_2l_3} \propto l^{-5/2}_{1}$,
while
the estimate for the local  bispectrum is $b^{loc}_{l_1l_2l_3} \sim l^{-2}_1$. 
So, our bispectrum is squeezed stronger than the seed local one.
This is a generic feature of models with intermediate Minkowskian stage.

Finally, we would like to address the issue of the observability of our bispectrum with the 
Planck data. To this end, 
we consider the Fisher matrix~\cite{Bartolo:2004if} 
\begin{equation}
\label{fisher}
F^{ij} \equiv \sum_{l_1 \le l_2 \le l_3} 
\frac{\langle a_{l_1m_1} a_{l_2m_2} a_{l_3m_3} \rangle^i \langle a_{l_1m_1} a_{l_2m_2} a_{l_3m_3}\rangle^{j}}{\sigma^2_{l_1l_2l_3}} \; .
\end{equation}
Here $\sigma^2_{l_1 l_2 l_3}$ is given by $\sigma^2_{l_1l_2l_3} =C_{l_1} C_{l_2} C_{l_3} D_{l_1l_2l_3}$ 
where $D_{l_1 l_2l_3}$ takes values 1, 2 and 6, when all $l$'s are different, 
two are the same and three are the same, respectively. 
Superscripts $i$, $j$ refer to 
particular sources of the non-Gaussianity, i.e. primordial physics, point 
sources, weak lensing, Sunyaev--Zel'dovich, 
emission etc. In what follows, we neglect all sources except for
primordial physics, so that the Fisher matrix reduces to a single 
element. Still, we stick to the standard notion of a 'matrix'.

To evaluate the sum over numbers 
$m_i$ in Eq.~\eqref{fisher}, we use the relation  
\begin{equation}
\nonumber 
\sum_{m_1m_2m_3} \left(
\begin{array}{ccc} 
l_1 & l_2 & l_3\\
m_1 & m_2 & m_3
\end{array}
\right)^2=1 \; .
\end{equation}
We estimate the angular power spectrum as $C_l \sim 6 \cdot 10^{-10} /l^2$, and
$3j$-symbols as
\begin{equation}
\nonumber
\left( \begin{array}{ccc} 
l_1 & l_2 & l_3\\
0 & 0 & 0
\end{array} \right) \sim \frac{1}{\sqrt{l_1l_2}} \; . 
\end{equation}
Now, using Eq.~\eqref{mainestimate} with $c_{l_1l_2l_3}
\sim 1$, and summing over multipole numbers $l_1$, $l_2$ and $l_3$, 
we obtain an estimate 
for the Fisher matrix
\begin{equation}
\label{fisherest} 
F \sim \frac{C^2}{1000\cdot z^4} \cdot \frac{{\bar l}^2}{l_0} \; .
\end{equation}
Here $\bar{l}$ and $l_0$ are the maximum and the minimum 
multipole numbers used in the analysis, respectively. 
Note that the dominant contributions to $F$ come from squeezed
configuratons with $l_3 \approx l_2 \sim \bar{l}$ and $l_1 \sim l_0$. 
We set $\bar{l}=2000$ and $l_0=10$ for estimates. 
The observation of the bispectrum in the CMB data is possible 
provided that the Fisher matrix is larger than unity, i.e. $F \gtrsim 1$. 
From Eq.~\eqref{fisherest}, we obtain phenomelogically interesting 
window for the parameter $z$
\begin{equation}
\nonumber 
l_0 \lesssim z \lesssim 5 \sqrt{|C|} \; .
\end{equation}
When writing the left 
inequality we recalled 
that the analysis of this Subsection is valid only for large enough 
values of the parameter $z$, i.e. $z \gg l_{min} \sim l_0 \sim 10$. Therefore, 
non-vanishing window for the 
parameter $z$ in the regime considered
requires 
the constant $C$ larger than about 100, which we find rather
contrived. 
Things are more optimistic 
in the situation with relatively short intermediate stage.

\subsection{Regime $1 \lesssim z \lesssim l_{min}$}

\begin{figure}[tb!]
\begin{center}
\includegraphics[width=0.44\columnwidth, angle=0]{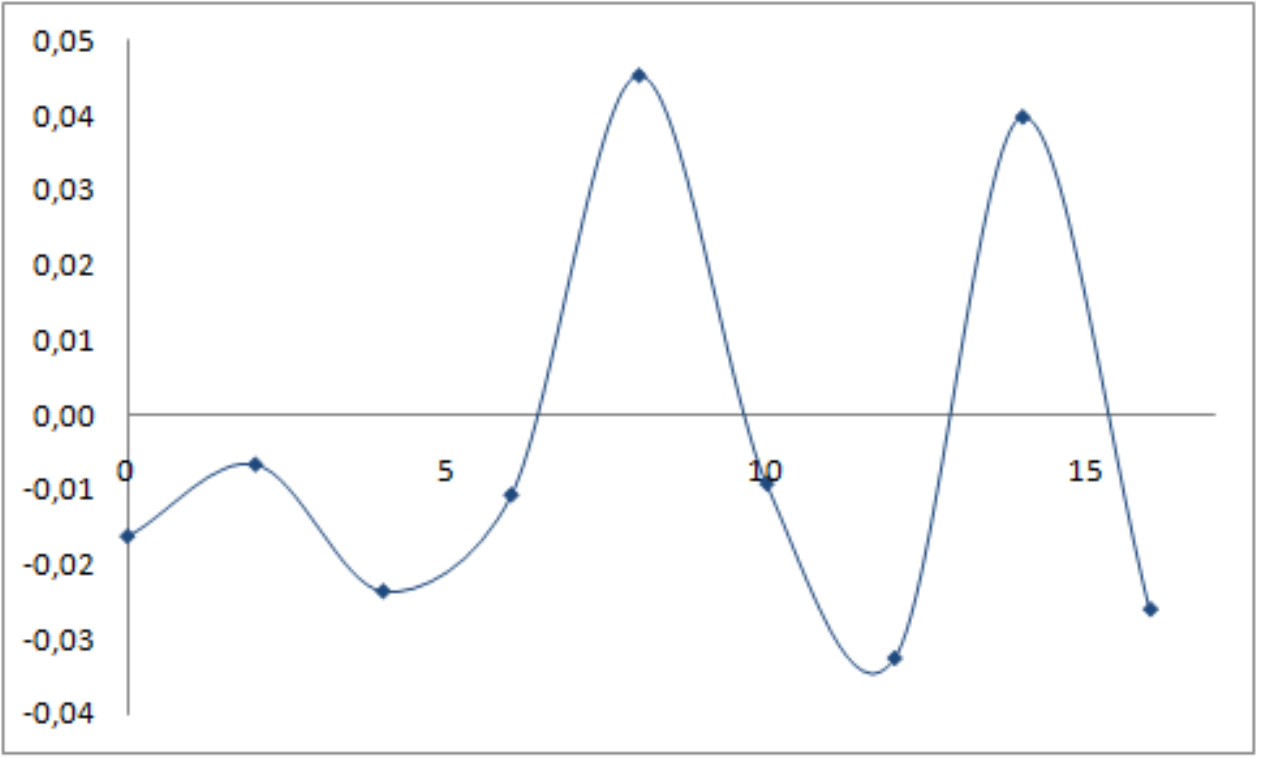}
\includegraphics[width=0.44\columnwidth, angle=0]{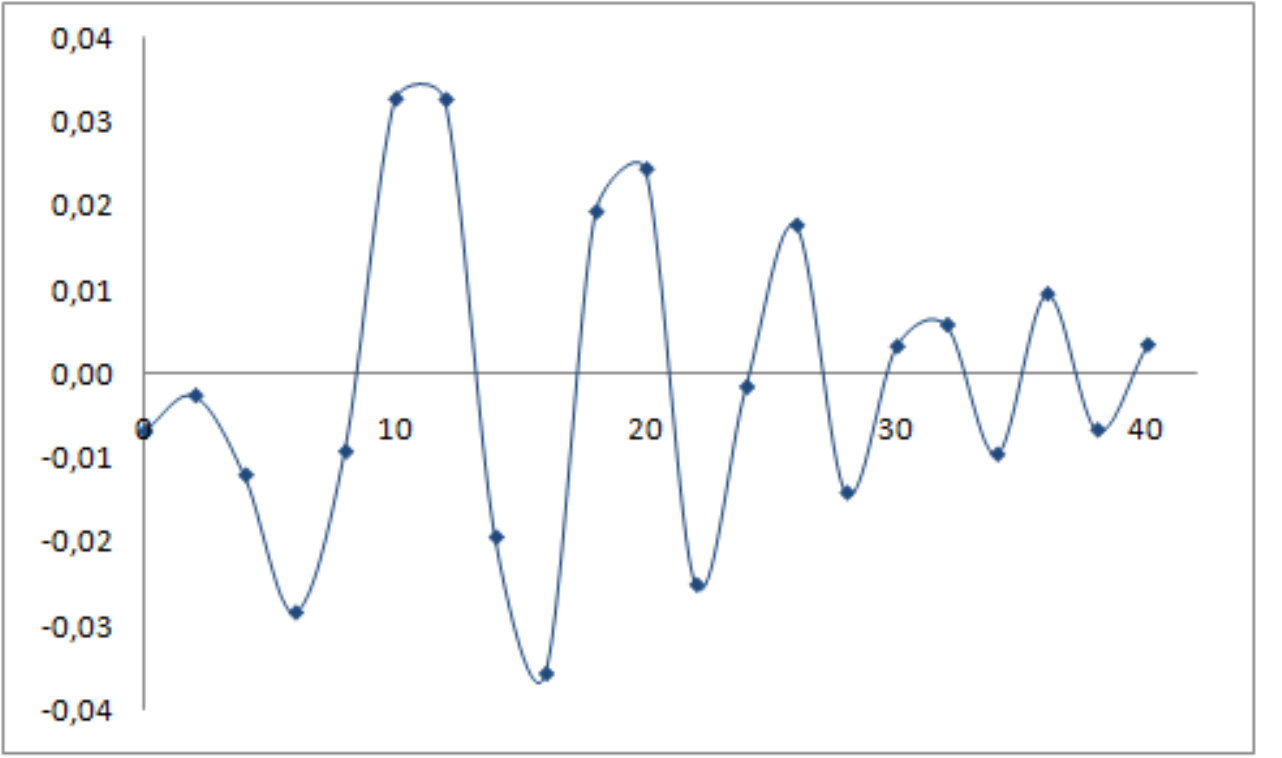}
\caption{Coefficients $c_{l_1l_2l_3}$ for the case of short intermediate stage 
as functions of the 
difference $\Delta l=l_1+l_2-l_3$ for $l_2=600$,  
$l_1=20$ (left) and $l_2=600$, $l_1=50$ (right). In both cases 
$z=10$.}\label{c2}
\end{center}
\end{figure}

For  $1 \lesssim z \lesssim l_{min}$, 
the estimate~\eqref{mainestimate} for the 
reduced bispectrum is replaced by the following one
\begin{equation}
\label{mainestimate2}
\begin{split}
|b_{l_1l_2l_3}| \sim  \frac{C}{30} \cdot \frac{{\cal P}^{3/2}_{\Phi}}{z}
\frac{\sqrt{l_3}}{l^{7/2}_1 \cdot l^{5/2}_2} |c_{l_1l_2l_3}| ,
\end{split}
\end{equation}
with coefficients $c_{l_1l_2l_3}$ given by 
\begin{equation}
\label{cl} 
\begin{split}
c_{l_1l_2l_3} &=l_1\Bigl( 1-\frac{\Delta l}{l_1} \Bigr) \int^{\infty}_1  \frac{du_1}{u^{7/2}_1} \cdot 
\frac{1}{[u^2_1-1]^{1/4}} \cdot \frac{1}{\Bigl[u^2_1-\Bigl(1-\frac{\Delta l}{l_1} \Bigr)^2\Bigr]^{1/2}} \times \\
& \times \Bigl ( \cos [l_1f(u_1)-(l_1-\Delta l) f(\bar{u}_2)]-\sin [l_1f(u_1)-(l_1-\Delta l) f(\bar{u}_2)]\Bigr) \times \\
& \times \sin \Bigl \{ \frac{1}{2} \frac{l_1l_2}{u_1l_3z} \Bigl [1+\Bigl(1-\frac{\Delta l}{l_1} \Bigr)^2 \Bigr]\Bigr \} 
\cdot \cos \Bigl \{\frac{l_1l_2}{u_1l_3z} \Bigl(1-\frac{\Delta l}{l_1} \Bigr) \Bigr \} \times \\
& \times \Bigl[1- \cot \Bigl \{ \frac{1}{2} \frac{l_1l_2}{u_1l_3z} 
\Bigl [1+\Bigl(1-\frac{\Delta l}{l_1} \Bigr)^2 \Bigr]\Bigr \} 
\cdot \tan 
\Bigl \{\frac{l_1l_2}{u_1l_3z} \Bigl(1-\frac{\Delta l}{l_1} \Bigr) \Bigr \} 
\cdot \frac{B^{l_1,1}_{l_2,-1;l_30}}{B^{l_10}_{l_20;l_30}}\Bigr] \; ,
\end{split}
\end{equation}
where $\Delta l >0$ ($c_{l_1l_2l_3}$ are again negligible for $\Delta l < 0$). 
Though this expression has more complicated structure, the qualitative behaviour of coefficients $c_{l_1l_2l_3}$ 
is fairly similar to that in Subsection 3.1. This can be readily seen from Fig.~\ref{c2}. Again we observe oscillations 
with the roughly constant order one amplitude. Equation~\eqref{mainestimate2} implies 
stronger squeezening as
compared to the case with the long intermediate stage. 

The estimate 
for the Fisher matrix changes accordingly,
\begin{equation}
\label{fisherest2}
F \sim \frac{C^2}{1000 \cdot z^2} \cdot \frac{{\bar l}^2}{l^3_0} 
\end{equation}
Substituting $\bar{l}=2000$ and $l_0=10$, we obtain a 
phenomenologically interesting window for the parameter $z$ 
\begin{equation}
\label{windowshort}
1 \lesssim z\lesssim |C| \; ,
\end{equation}
existing now for $C \gtrsim 1$, which is, 
of course, welcome from the viewpoint of microscopic physics. For even smaller values $|C| \lesssim 1$, however, 
the bispectrum tends to be of the non-observable size. 
It is tempting to relax this constraint by extrapolating the analysis down to $l_0=2$. 
In that case the
phenomenologically interesting window could be widened by about an order of magnitude. 
In turn, this would imply an 
order of magnitude increase in the sensitivity to the parameter $C$. 
Recall, however, that we are working in the large-$l$ approximation. 
Therefore, our calculations and, in particular, the estimate \eqref{fisherest2}
for the Fisher matrix, are not justified for values of $l_0$ as small as 2.

.

\section{Conclusion}

Let us summarize the effects due to the 
presence of the intermediate Minkowskian evolution: 
\begin{itemize} 
 \item The bispectrum is suppressed by the duration of the intermediate stage. 
We observed that the bispectrum behaves as $\eta^2_0/|\eta_*|^2$ 
for long enough duration $|\eta_*|$ and as $\eta_0/ |\eta_*|$ provided that the duration 
is relatively  short;  
 \item The bispectrum vanishes for $l_1+l_2 < l_3$ and
undergoes oscillations as function
of $\Delta l=l_1+l_2-l_3 > 0$ (our convention is
$l_1 \leq l_2 \leq l_3$). 
The origin of oscillations can be traced back to oscillations  in the
CMB transfer 
functions relating the  temperature fluctuation to the Newtonian potential. 
Normally, these oscillations are averaged out upon integrating 
over directions of momenta ${\bf k}$. In our case this does not happen 
because the bispectrum of scalar perturbations
is saturated in the 
region where ${\bf k}_2$ is almost parallel to ${\bf k}_1$.

\item Oscillations are characterized by roughly constant amplitude. 
This fact is quite non-trivial. 
Indeed, direct translation of the relation $k_3 \approx k_1+k_2$ 
into the harmonic space naively implies the preference
of flattened trinagle configurations 
obeying the relation $l_3 \approx l_1 +l_2$. This tendency is indeed there, 
as is clearly seen from Fig.~\ref{fa}. Unexpectedly, the opposite tendency 
is also present: there are cancellations occuring for 
configurations obeying flattened triangle relation. 
The interplay between the 
two tendencies is such that the amplitude gets smoothened, and we do not observe any preference of
flattened over non-flattened configurations or vice versa. This 
feature is important for discriminating the models we discussed in this
paper from inflationary scenarios, some of 
which 
predict the oscillatory behaviour of the bispectrum analogous to ours. 
For example, 
shapes somewhat
similar to one in 
Eq.~\eqref{bimoment} are considered in 
Refs.~\cite{Holman:2007na,Chialva:2011hc,Meerburg:2009ys, Agullo:2010ws, Chen:2010bka}.
 These are, however, characterized by the amplitude strongly peaked 
in the flattened triangle limit. 

 \item In the situation with the long intermediate stage, there is squeezing of the 
CMB bispectrum as compared to the case 
where the intermediate stage is absent. Squeezing is 
particularly strong if the duration of the 
stage is smaller than the minimum 
multipole number $l_1$. 
\end{itemize}



We also discussed the sensitivity of the Planck experiment to
the bispectrum predicted. We concluded that there is a window for the duration 
of the intermediate stage, though not particularly wide, 
where the non-Gaussianity can be of the observable size. 
Note that the signal of interest, if detected in the CMB data, 
would not imply the 
specific early Universe model, but rather indicate the presence of the 
long Minkowskian evolution before the hot era. 
This could be of particular importance from the viewpoint 
of alternatives to inflation, e.g. ekpyrotic scenarios and Galilean Genesis. 




\section*{Acknowledgments}

We thank Diego Chialva, Dmitry Kirpichnikov, Maxim Libanov, Grigory
Rubtsov and especially Denis Karateev for valuable comments and
discussions. This work is supported in part by RFBR grant 12-02-00653
(S.M. and V.R.), the grant of the President of the Russian Federation
NS-2835.2014.2 (S.M. and V.R.), the Dynasty Foundation (S.M.) and by
the Belgian Science Policy IAP VII/37 (S.R.). S.R. is also indebted to
Lund University for warm hospitality during some stage of this work.

\section*{Appendix A. Derivation of the explicit term in Eq.~\eqref{bileading}}

In this Appendix, we comment on the details of the derivation of the 
formula~\eqref{bileading} 
valid for $z \gg l_1$.
As discussed in Subsection 2.2, the contribution of the order $1/z$ vanishes in that case. 
To calculate the bispectrum to the order $1/z^2$, we expand slowly varying functions 
of angles $(\theta_1, \phi_1)$ and $(\theta_3, \phi_3)$ up to the second order, i.e.  
\begin{equation}
\nonumber
\sin \theta_1 =\sin \theta_2 +\cos \theta_2 \delta \theta_1 -\frac{1}{2} \sin \theta_2 \delta \theta^2_1 \; ,
\end{equation}
\begin{equation}
\nonumber
\begin{split}
Y^{*}_{l_1 m_1} (\theta_1, \phi_1) &=Y^{*}_{l_1, m_1} (\theta_2, \phi_2) +\frac{\partial Y^{*}_{l_1 m_1}}{\partial \theta}
\delta \theta_1 +\frac{\partial Y^{*}_{l_1, m_1}}{\partial \phi} \delta \phi_1 +\\
& +\frac{1}{2} \frac{\partial^2 Y^{*}_{l_1 m_1}}{\partial \theta^2} \delta \theta^2_1+
\frac{\partial^2 Y^{*}_{l_1 m_1}}{\partial \phi \partial \theta}  \delta \theta_1 \delta \phi_1 + 
\frac{1}{2}\frac{\partial^2 Y^{*}_{l_1 m_1}}{\partial \phi^2}  \delta \phi^2_1 \; ,
\end{split} 
\end{equation}
and
\begin{equation}
\nonumber
\begin{split}
Y^{*}_{l_3 m_3} (\theta_3, \phi_3) &=Y^{*}_{l_3, m_3} (\bar{\theta}_3, \bar{\phi}_3) +
\frac{\partial Y^{*}_{l_3 m_3}}{\partial \theta_3}
\delta \theta_3 +\frac{\partial Y^{*}_{l_3, m_3}}{\partial \phi_3} \delta \phi_3 +\\
& +\frac{1}{2} \frac{\partial^2 Y^{*}_{l_3 m_3}}{\partial \theta^2_3} \delta \theta^2_3+ 
\frac{\partial^2 Y^{*}_{l_3 m_3}}{\partial \theta_3 \partial \phi_3}  \delta \theta_3 \delta \phi_3 + 
\frac{1}{2} \frac{\partial^2 Y^{*}_{l_3 m_3}}{\partial \phi^2_3} \delta \phi^2_3\; . 
\end{split} 
\end{equation} 
Here $\bar{\theta}_3=\pi -\theta_2$ and $\bar{\phi}_3 =\pi +\phi_2$, 
derivatives of the spherical 
harmonic $Y^{*}_{l_3m_3}$ with respect to $\theta_3$ and $\phi_3$ are taken at points 
$\bar{\theta}_3$ and $\bar{\phi}_3$, respectively. Variations of 
angles $\theta_3$ and $\phi_3$ are not independent but are
related to variations of angles $\theta_1$ and $\phi_1$,   
\begin{equation}
\label{theta3}
\delta \theta_3 =\frac{1}{4} \frac{y_1 y_2}{(y_1 +y_2)y_3} \sin 2\theta_2  (\delta \phi_1)^2 -
\frac{y_1}{y_3} \delta \theta_1 
\end{equation} 
and
\begin{equation}
\label{phi3}
\delta \phi_3 =\frac{y_1}{y_3 \sin^2 \theta_2} \left(\sin^2 \theta_2 -\frac{y_1}{y_3} \sin \theta_2 \cos \theta_2 
\delta \theta_1 +\sin \theta_2 \cos \theta_2 \delta \theta_1 \right) \delta \phi_1 \; .
\end{equation} 
We also expand the cosine,
\begin{equation} 
\label{cosinefour}
\begin{split}
&\cos ([y_1 +y_2 -|{\bf y}_1 +{\bf y}_2|]z)=\cos \Bigl(N[(\delta \theta_1)^2 +\sin^2 
\theta_2 (\delta \phi_1)^2] \Bigr) 
\Bigl(1-\frac{1}{8}N^2 \sin^2 2\theta_2 (\delta \theta_1)^2 (\delta \phi_1)^4 \Bigr)-\\ 
&-\sin \Bigl( N[(\delta \theta_1)^2 +\sin^2 \theta_2 (\delta \phi_1)^2]\Bigr) N\Bigl[ -
\frac{\delta \theta^4_1}{12} -\sin^2 \theta_2 \frac{\delta \phi^4_1}{12}-\\ 
&-\frac{1}{2} \sin^2 \theta_2 \delta \theta^2_1 \delta \phi^2_1 +\frac{1}{2} \sin 2 \theta_2 \delta \theta_1 \delta \phi^2_1 
+\frac{y_1 y_2}{4(y_1 +y_2)^2}\Bigl( \delta \theta^4_1 +2 \sin^2 \theta_2 \delta \theta^2_1 \delta \phi^2_1 
+\sin^4 \theta_2 \delta \phi^4_1\Bigr)\Bigr] \; .
\end{split} 
\end{equation} 
Combining these expressions and using integrals
\begin{equation}
\nonumber
\int^{+\infty}_{-\infty} \cos Nx^2 dx =\int^{+\infty}_{-\infty} \sin N x^2 dx =\frac{1}{\sqrt{N}}\sqrt{\frac{\pi}{2}} \; ,
\end{equation}
\begin{equation}
\nonumber
\int^{+\infty}_{-\infty} x^2\cos Nx^2 dx=-\int^{+\infty}_{-\infty} x^2 \sin Nx^2 dx=-\frac{\sqrt{2\pi}}{4N^{3/2}} \; ,
\end{equation}
\begin{equation}
\nonumber
\int^{+\infty}_{-\infty} x^4 \cos Nx^2 dx =\int^{+\infty}_{-\infty} x^4 \sin Nx^4 dx =-\frac{3\sqrt{2\pi}}{8N^{5/2}} \; , 
\end{equation}
we obtain for the leading contribution, 
\begin{equation}
\label{inter}
\begin{split}
\langle a_{l_1m_1} a_{l_2m_2} a_{l_3 m_3} \rangle_0 &=\frac{\pi {\cal P}^{3/2}_{\Phi}}{4z^2} \int dy_1 dy_2 (y_1+y_2)^2\Delta_{l_1} (y_1) 
\Delta_{l_2} (y_2) \Delta_{l_3} (y_1+y_2) \times \\
& \times A(y_1, y_2, y_1+y_2,z) \times \\
& \times \Bigl( B^{l_1, m_1}_{l_2, m_2; l_3, m_3} \Bigl[ l_1(l_1+1)+\frac{y^2_1}{(y_1+y_2)^2}l_3(l_3+1) 
+\frac{2y_1y_2}{(y_1+y_2)^2}\Bigr]+\\
&+\frac{2y_1}{y_1+y_2}F^{l_1m_1}_{l_2,m_2; l_3, m_3}\Bigr) \; .
\end{split}
\end{equation}
Here coefficients $B^{l_1m_1}_{l_2m_2;l_3m_3}$ are given by Eq.~\eqref{B1}, while quantities 
$F^{l_1m_1}_{l_2m_2;l_3m_3}$ are defined by 
\begin{equation}
\nonumber
F^{l_1m_1}_{l_2m_2; l_3m_3} =-\int d\theta d\phi \sin \theta Y^{*}_{l_2m_2} 
\left(\frac{\partial Y^{*}_{l_1m_1}}{\partial \theta} \frac{\partial Y^{*}_{l_3m_3}}{\partial \theta} 
+\frac{1}{\sin^2 \theta}\frac{\partial Y^{*}_{l_1m_1}}{\partial \phi} \frac{\partial Y^{*}_{l_3m_3}}{\partial \phi} \right)\; .
\end{equation}
When deriving Eq.~\eqref{inter}, we used the master equation for spherical harmonics, 
\begin{equation}
\label{master}
\frac{\partial^2 Y_{lm}}{\partial \theta^2} +\mbox{ctg} \theta \cdot \frac{\partial Y_{lm}}{\partial \theta} -\frac{m^2}{\sin^2 \theta} 
Y_{lm} +l(l+1) Y_{lm} =0 \; .
\end{equation}
Note that Eq.~\eqref{inter} does not look symmetric under 
interchange of multipole numbers $(l_1,m_1) \leftrightarrow (l_2,m_2)$. To obtain the bispectrum in the symmetric form 
we get rid of derivatives of the spherical harmonic $Y^{*}_{l_3m_3} (\theta, \phi)$ by performing integration by parts 
in Eq.~\eqref{inter}. We also
make use of the master equation~\eqref{master} to get rid of 
the term $l_3(l_3+1)Y^{*}_{l_3,m_3}$. Again performing the integration by parts, we obtain 
\begin{equation}
\label{bileading1}
\begin{split} 
\langle a_{l_1 m_1} a_{l_2 m_2} a_{l_3 m_3} \rangle_0 &=i^{l_1+l_2 +l_3} {\cal P}^{3/2}_{\Phi} 
\frac{\pi}{4z^2} \int d y_1 d y_2 \Delta_{l_1} 
(y_1) \Delta_{l_2} (y_2) \Delta_{l_3} (y_1+y_2) \times \\
& \times A(y_1, y_2, y_1+y_2) \Bigl \{ 2y_1 y_2 D^{l_1,m_1}_{l_2, m_2; l_3, m_3}+\\
& +B^{l_1, m_1}_{l_2, m_2; l_3, m_3} \Bigl[y^2_2l_1(l_1+1) +y^2_1l_2(l_2+1)+2y_1 y_2 \Bigr] \Bigr \} 
+(l_1,~l_2 \leftrightarrow l_3) \; ,
\end{split}
\end{equation}

Coefficients $D^{l_1, m_1}_{l_2, m_2; l_3, m_3}$ are defined by
\begin{equation}
\label{D}
D^{l_1 m_1}_{l_2 m_2; l_3 m_3} =\int d \phi d \theta \sin \theta \left[ \frac{\partial Y^{*}_{l_1 m_1}}{\partial \theta} 
\frac{\partial Y^{*}_{l_2 m_2}}{\partial \theta} +\frac{1}{\sin^2 \theta} \cdot 
\frac{\partial Y^{*}_{l_1 m_1}}{\partial \phi} 
\frac{\partial Y^{*}_{l_2 m_2}}{\partial \phi}\right] Y^{*}_{l_3 m_3} (\pi-\theta, \pi +\phi) \; .
\end{equation}
Note that Eq.~\eqref{bileading1} is symmetric under the 
interchange of multipole numbers $(l_1,m_1) \leftrightarrow (l_2, m_2)$. 

It is worth noting that Eq.~\eqref{bileading1} represents statistically isotropic bispectrum. 
Namely, the dependence of the latter on numbers $m_i$ obeys the condition~\cite{Bartolo:2004if} 
\begin{equation}
\label{property}
\langle a_{l_1m_1} a_{l_2m_2} a_{l_3m_3} \rangle = \sum_{m'_1m'_2m'_3} \langle a_{l_1m'_1}a_{l_2m'_2} a_{l_3m'_3} \rangle 
D^{(l_1)}_{m'_1m_1} D^{(l_2)}_{m'_2m_2} D^{(l_3)}_{m'_3m_3} \; .
\end{equation}
Here $\langle a_{l_1m'_1} a_{l_2m'_2} a_{l_3m'_3} \rangle $ denotes the bispectrum obtained by the rotation 
of the 
celestial sphere by Euler angles $\alpha$, $\beta$ and $\gamma$; 
coefficients $D^{(l)}_{m'm}=D^{(l)}_{m'm} (\alpha, \beta, \gamma)$ are 
matrix elements of the rotation matrix $D (\alpha, \beta, \gamma)$, i.e.
\begin{equation}
\nonumber 
Y_{lm'} (\theta', \phi')=\sum^{l}_{m=-l} D^{(l)}_{m'm} Y_{lm} (\theta,
\phi) \; .
\end{equation}
Equation~\eqref{property} is satisfied, since this is the property of coefficients 
$B^{l_1m_1}_{l_2m_2;l_3m_3}$ and $D^{l_1m_1}_{l_2m_2;l_3m_3}$. Indeed, we write the former as in Eq.~\eqref{B} 
and then recall the well known property of the $3j$-symbols~\cite{Bartolo:2004if} 
\begin{equation}
\nonumber 
\left( \begin{array}{ccc} 
l_1 & l_2 & l_3\\
m_1 & m_2 & m_3
\end{array} \right) = \sum_{m'_1m'_2m'_3} \left( \begin{array}{ccc} 
l_1 & l_2 & l_3\\
m'_1 & m'_2 & m'_3
\end{array} \right) 
D^{(l_1)}_{m'_1m_1} D^{(l_2)}_{m'_2m_2} D^{(l_3)}_{m'_3m_3} \; .
\end{equation}
Clearly, the derivative structure in 
the integrand of Eq.~\eqref{D} also preserves statistical isotropy\footnote{We 
thank D.~Karateev for the numerical check of this somewhat heuristic statement.}. 
This 
determines the dependence of coefficients 
$D^{l_1m_1}_{l_2m_2;l_3m_3}$ on  $m_i$:
\begin{equation}
\label{dstat}
D^{l_1m_1}_{l_2m_2;l_3m_3} = \left( \begin{array}{ccc} 
l_1 & l_2 & l_3\\
m_1 & m_2 & m_3
\end{array} \right) D_{l_1l_2l_3}\; ;
\end{equation}
coefficients $D_{l_1l_2l_3}$ introduced here are independent of numbers $m_i$. We use Eq.~\eqref{dstat} 
to derive coefficients $D^{l_1m_1}_{l_2m_2;l_3m_3}$ in terms of $3j$-symbols. 
Namely, we evaluate the integral~\eqref{D} for $m_1=m_2=m_3=0$.
To this end, we use the following relations 
for derivatives of spherical harmonics, 
\begin{equation}
\nonumber
\frac{\partial Y_{l0} (\theta)}{\partial \theta}=
\sqrt{l(l+1)} Y_{1,1} (\theta, \phi) e^{-i\phi} =-\sqrt{l(l+1)} Y_{1,-1} (\theta, \phi)e^{i\phi} \; .
\end{equation}
Substituting them into Eq.~\eqref{D}, using Eq.~\eqref{B} with $m_1=-m_2=1$ and $m_3=0$
and 
Eq.~\eqref{dstat}, we obtain coefficients $D_{l_1l_2l_3}$. Substituting the latter back into 
Eq.~\eqref{dstat}, but with arbitrary numbers $m_i$, we derive the expression of interest, 
\begin{equation}
\label{Df}
D^{l_1m_1}_{l_2m_2;l_3 m_3} =-\sqrt{l_1(l_1+1)}\sqrt{l_2(l_2+1)} B^{l_1,1}_{l_2,-1;l_3,0} (B^{l_10}_{l_20;l_30})^{-1} 
B^{l_1m_1}_{l_2m_2;l_3m_3} \; .
\end{equation} 
Finally, substituting these into Eq~\eqref{bileading1}, we obtain the term written explicitly in Eq.~\eqref{bileading}.

\section*{Appendix B. Comments on terms with permutations in Eq.~\eqref{bileading}}
In this Appendix, we comment on terms with permutations of multipole 
numbers in the last line of Eq.~\eqref{bileading}. We 
consider the case $l_3 \sim l_2 \gg l_1 \gg 1$ and
start with the term with interchange
$(l_1,m_1) \leftrightarrow (l_3,m_3)$, 
\begin{equation}
\label{perm}
\begin{split}
\langle a_{l_1m_1} a_{l_2m_2} a_{l_3m_3} \rangle^{l_1 \leftrightarrow l_3} &=i^{l_1+l_2+l_3} 
\cdot \frac{\pi {\cal P}^{3/2}_{\Phi}}{4z^2} 
\cdot \int dy_2 dy_3 \Delta_{l_1} (y_2+y_3) \Delta_{l_2} (y_2) \Delta_{l_3} (y_3)\times \\ 
& \times A(y_2+y_3,y_2,y_3) \cdot B^{l_1m_1}_{l_2m_2;l_3m_3} \cdot 
\Bigl \{ y^2_2l_3(l_3+1) +y^2_3l_2(l_2+1)  -\\ 
&-2y_2y_3 \sqrt{l_2(l_2+1)} \sqrt{l_3(l_3+1)} 
B^{l_10}_{l_2,-1;l_3,1} \Bigl(B^{l_10}_{l_20;l_30} \Bigr)^{-1} +2y_2y_3\Bigr \} \; .
\end{split}
\end{equation}
Due to the properties of the transfer functions, the integration range here
is effectively $y_2 \geq l_2 +1/2$, $y_3 \geq l_3 +1/2$. 
Naively, Eq.~\eqref{perm} gives the contribution 
enhanced by a factor $\sim l^3_3/l^3_1$ as compared 
to one written explicitly in Eq.~\eqref{bileading}. 
Note, however, that generically shape functions $A(y_1,y_2,y_3)$ are suppressed 
for configurations with $y_1 \approx y_2+y_3 \gg l_1$ and $y_2,~y_3 \sim l_3$ as compared to configurations 
with $y_1 \sim l_1$, $y_2 \sim l_2 \sim l_3$ and $y_3 \approx y_1+y_2 \sim l_3$. 
In particular, for the standard local shape bispectrum as in Eq.~\eqref{local} one has
\begin{equation}
\nonumber
A^{loc}(y_2+y_3,y_2,y_3) \left. \right|_{y_2,~y_3 \sim l_3} 
\sim \frac{l^3_1}{l^3_3} \cdot A^{loc}(y_1,y_2,y_1+y_2) \left. \right|^{y_1 \sim l_1}_{y_2 \sim l_2 \sim l_3} \; .
\end{equation}
Furthermore, 
there is the suppression by a factor $\sim l_1/l_3$ in the transfer function 
$\Delta_{l_1} (y_2+y_3)$, i.e. 
\begin{equation}
\nonumber
\Delta_{l_1} (y_2+y_3) \left. \right|_{y_2 \sim y_3 \sim l_3} \sim \frac{l_1}{l_3} 
\cdot \Delta_{l_1} (y_1) \left. \right|_{y_1 \sim l_1} \; .
\end{equation} 
This follows from the approximate form of the spherical Bessel function~\eqref{spherBessel}. 
Finally, for configurations $y_1 \approx y_2+y_3$, there is no saddle point in the product 
of three transfer functions. This gives 
extra suppression by a factor $l^{3/2}_1/l^2_3$ as 
compared to the explicit term in Eq.~\eqref{bileading}. In total, we have the estimate for the contribution 
to the reduced bispectrum due to the
permutation $(l_1,m_1) \leftrightarrow (l_3,m_3)$, 
\begin{equation}
\label{just1}
b^{l_1 \leftrightarrow l_3}_{l_1l_2l_3} \sim b_{l_1l_2l_3} \cdot \frac{l^{5/2}_1}{l^3_3} \; .
\end{equation}
Two  remarks are in order. First, we used the local type of 
the shape function $A(y_1,y_1,y_3)$, Eq.~\eqref{local}, 
in our estimate. For other types of shape functions, the 
suppression is generically milder, but it is
still there. In particular, the suppression is of the order 
$\sqrt{l_1}/{l_3}$ for the equilateral type of the 
bispectrum generated before the intermediate stage. 
Second, we assumed long Minkowskian stage, $z\gg l_3$. 
For smaller values of the parameter $z$, i.e. $l_1 \ll z \lesssim l_3$, the expression 
in the r.h.s. of Eq.~\eqref{perm} should be replaced by Eq.~\eqref{bishort} (with multipole numbers $(l_1,m_1)$ and $(l_3,m_3)$ 
permuted). Clearly, additional suppression by a factor $z/l_3$ takes place in that case. The latter 
becomes of the order $l_1/l_3$ for even 
smaller duration of the intermediate stage, i.e. for $z \lesssim l_1$.  

Finally, let us comment on the term with permuted
multipole numbers $(l_2,m_2)$ and $(l_3,m_3)$ 
in Eq.~\eqref{bileading}. It is also suppressed, albeit
mildly, as compared to one written explicitly in Eq.~\eqref{bileading}. 
The reason is that the product of the
three transfer functions is a rapidly oscillating function for all relevant values of variables $y_i$. 
That is, it does not have the saddle point akin to the case discussed 
in the main text. 
As a result, we have an estimate for the term of interest, 
\begin{equation}
\label{just2}
b^{l_2 \leftrightarrow l_3}_{l_1l_2l_3} \sim \frac{1}{\sqrt l_1} b_{l_1l_2l_3} \; .
\end{equation}
This estimate works for all relevant values of the parameter $z$.

Estimates~\eqref{just1} and~\eqref{just2} justify neglecting terms 
with permutations in Section~3.


\section*{Appendix C. Details of saddle-point calculation in the generic case}

Here we comment on the derivation of Eq.~\eqref{bishort}. To simplify calculations, we use the same 
trick as in Appendix~A and 
calculate the bispectrum for $m_1=m_2=m_3=0$.
Using the property of 
statistical isotropy, we multiply the result by $B^{l_1m_1}_{l_2m_2;l_3m_3}/B^{l_10}_{l_20;l_30}$ and obtain the 
bispectrum valid for arbitrary $m_i$. 

Using approximate formula~\eqref{spherap}  with
$m_1=m_2=m_3=0$, we write for the 
product of three spherical harmonics and the cosine
\begin{equation}
\nonumber
\begin{split}
&Y_{l_10} (\theta_1) Y_{l_20} (\theta_2) Y_{l_30} (\theta_3)
  \cos \Bigl(N [\delta \theta^2_1 +\sin^2 \theta_2 \delta \phi^2_1 ] \Bigr) =\\
& =\frac{1}{4} Y_{l_10} (\theta_2) Y_{l_20} (\theta_2) Y_{l_30} (\bar{\theta}_3)  \times \\
& \times \Bigl[ \cos \Bigl(l_1 \delta \theta_1 +l_3 \delta \theta_3 
+N [\delta \theta^2_1+\sin^2 \theta_2\delta \phi^2_1] \Bigr) +\cos \Bigl(l_1 \delta \theta_1 +l_3 \delta \theta_3- 
N [\delta \theta^2_1+\sin^2 \theta_2\delta \phi^2_1] \Bigr)+\\
&+\cos \Bigl(l_1 \delta \theta_1 -l_3 \delta \theta_3+ 
N [\delta \theta^2_1+\sin^2 \theta_2 \delta \phi^2_1] \Bigr)+\cos \Bigl(l_3 \delta \theta_3+ 
N [\delta \theta^2_1+\sin^2 \theta_2 \delta \phi^2_1] -l_1 \delta \theta_1\Bigr) \Bigr]+\\
&+\frac{1}{4l_1l_3} \frac{\partial Y_{l_10} (\theta_2)}{\partial \theta_2} \cdot Y_{l_20} (\theta_2) \cdot \frac{\partial Y_{l_30} (\bar{\theta}_3)}{\partial \theta_2} \times \\
& \times \Bigl[ \cos \Bigl(l_1 \delta \theta_1 +l_3 \delta \theta_3 
+N [\delta \theta^2_1+\sin^2 \theta_2\delta \phi^2_1] \Bigr) +\cos \Bigl(l_1 \delta \theta_1 +l_3 \delta \theta_3- 
N [\delta \theta^2_1+\sin^2 \theta_2\delta \phi^2_1] \Bigr)-\\
&-\cos \Bigl(l_1 \delta \theta_1 -l_3 \delta \theta_3+ 
N [\delta \theta^2_1+\sin^2 \theta_2 \delta \phi^2_1] \Bigr)-\cos \Bigl(+l_1 \delta \theta_1-l_3 \delta \theta_3-N [\delta \theta^2_1+\sin^2 \theta_2 \delta \phi^2_1] \Bigr) \Bigr]+\\
&+... \; .
\end{split}
\end{equation}
Here dots 
stand for the combination of terms which is 
antisymmetric under 
$\delta \theta_1 \to -\delta \theta_1$ and vanishes when integrated over $\theta_1$. 
Saddle point of the first cosine in the third line is at
\begin{equation}
\nonumber
\delta \bar{\theta}_1 =\frac{y_1(l_3-l_1)-l_1y_2}{y_1y_2z} \; .
\end{equation}
Similar expressions hold for saddle points of other cosines. 
Now, performing the integration in the vicinity of these saddle points, we obtain
\begin{equation}
\nonumber
\begin{split}
\langle a_{l_10} a_{l_20} a_{l_30} \rangle &=i^{l_1+l_2+l_3} \cdot \frac{\pi{\cal P}^{3/2}_{\Phi}}{4}  \cdot
B^{l_10}_{l_20;l_30} \times \\ 
& \times \int  dy_1 dy_2 \cdot y^2_1 \cdot y^2_2 \cdot \Delta_{l_1} (y_1) \Delta_{l_2} (y_2) \Delta_{l_3} (y_1+y_2) 
 \cdot A(y_1,y_2,y_1+y_2) \times \\
& \times \frac{1}{N} \cdot \sin \Bigl \{\frac{1}{4N} \Bigl(l^2_1+\frac{y^2_1l^2_3}{(y_1+y_2)^2} \Bigr) \Bigr \} \cdot \cos 
\Bigl \{\frac{1}{2N} \cdot \frac{y_1l_1l_3}{y_1+y_2} \Bigr \} \times \\
& \times \Bigl[1-
\cot \Bigl \{\frac{1}{4N} \Bigl(l^2_1+\frac{y^2_1l^2_3}{(y_1+y_2)^2} \Bigr) \Bigr \} \cdot 
\tan \Bigl \{\frac{1}{2N} \cdot \frac{y_1l_1l_3}{y_1+y_2} \Bigr \} \times \\ 
& \times \Bigl(\frac{l_1}{l_3} +\frac{l_2}{l_3} \cdot 
B^{l_1,1}_{l_2,-1;l_30} (B^{l_10}_{l_20;l_30})^{-1} \Bigr)\Bigr]\\
&  + (l_1,m_1\leftrightarrow l_3,m_3) + (l_2,m_2 \leftrightarrow l_3,m_3)\; .
\end{split}
\end{equation}
Finally, integrating by parts and multiplying the result by 
$B^{l_1m_1}_{l_2m_2;l_3m_3}/B^{l_10}_{l_20;l_30}$, we obtain the expression~\eqref{bishort} from the main 
body of the text. 

\section*{Appendix D. Comments on the derivation of the estimates~\eqref{mainestimate} and~\eqref{mainestimate2}}
In this Appendix, we comment on the 
derivation of the estimates~\eqref{mainestimate} and~\eqref{mainestimate2}. 
For this purpose, we take the integrals over the variables $y_1$ and $y_2$ present in Eqs.~\eqref{bileading} and~\eqref{bishort}. 
With the simplification~\eqref{replace}, transfer functions $\Delta_l (y)$ become proportional to 
spherical Bessel functions $j_l (y)$. An excellent approximation to spherical Bessel functions in the regime $y > l+1/2$ 
is given by
\begin{equation}
\label{spherBessel}
j_{l} (y) =\frac{1}{\left(l+\frac{1}{2} \right)} \frac{1}{\sqrt{u}}\frac{1}{[u^2-1]^{1/4}} 
\cos \left[\left( l+\frac{1}{2}\right)f(u)-\frac{\pi}{4}\right] \; , 
\end{equation}
where 
\begin{equation}
\nonumber
u=\frac{y}{l+1/2}
\end{equation}
and 
\begin{equation}
\nonumber
f(u) =\sqrt{u^2-1} -\mbox{arccos} \left( \frac{1}{u}\right) \; .
\end{equation}
In the region $y<l+1/2$, i.e. for $u<1$, the 
spherical Bessel function is exponentially suppressed and 
we can safely neglect the contribution coming from this region. We evaluate  
the integral in Eq.~\eqref{bileading} or Eq.~\eqref{bishort} 
over the variable $u_2$ first. 
Once again, we employ the saddle point 
technique. This is legitimate because of 
the large-$l$ factor in the argument of cosine in Eq.~\eqref{spherBessel}. 
In the integrand, the product of spherical Bessel functions can be written as follows   
\begin{equation}
\label{prodbessel}
  j_{l_2} (y_2) j_{l_3} (y_3) \approx
 \frac{1}{2l_2l_3[u^2_2 \cdot u^2_3 \cdot(u^2_2-1) \cdot (u^2_3-1)]^{1/4}}\cos \left[l_2 f(u_2) -l_3 f(u_3) \right] \; ,
\end{equation}
where 
\begin{equation}
\nonumber
u_3 =\frac{l_1}{l_3} u_1 +\frac{l_2}{l_3}u_2 \; .
\end{equation}
and we dropped the term 
with the second cosine which is a rapidly oscillating function 
for all values of $u_2$. 
The saddle point of the expression~\eqref{prodbessel} is at  
\begin{equation}
\nonumber
\bar{u}_2 =\bar{u}_3=\frac{l_1}{l_3-l_2} u_1 \; .
\end{equation}
We expand the argument of the cosine in Eq.~\eqref{prodbessel} up to the quadratic 
term in $\Delta u_2 \equiv u_2 -\bar{u}_2$
\begin{equation}
\nonumber
[l_2 f(u_2)-l_3 f(u_3)]=[l_2 f(\bar{u}_2)-l_3 f(\bar{u}_3)] +\frac{l_2 (l_3-l_2)^4 }{l^3_1 l_3u^2_1\sqrt{u^2_1-\frac{(l_3-l_2)^2}{l^2_1}}} \frac{\Delta u^2_2}{2} \; .
\end{equation}  
The integral over $u_2$ is now evaluated 
in a straightforward manner. Final estimates for the bispectrum 
are presented in the main body of the article 
in terms of the remaining integral over the variable $u_1$. 
They are given by Eqs.~\eqref{mainestimate},~\eqref{clll} and~\eqref{f} for the case of long intermediate 
stage and by Eqs.~\eqref{mainestimate2} and~\eqref{cl} for sufficiently short intermediate stage.

\end{document}